\documentclass[apj]{emulateapj}
\usepackage{amsmath}

\begin{document}


\title{X-ray Sources in the Dwarf Spheroidal Galaxy Draco}


\author{E. Sonbas\altaffilmark{1,2}, B. Rangelov\altaffilmark{2}, O. Kargaltsev\altaffilmark{2}, K.S. Dhuga\altaffilmark{2}, J. Hare\altaffilmark{2}, and I. Volkov\altaffilmark{2,3}}
\affil{$^1$University of Adiyaman, Department of Physics, 02040 Adiyaman, Turkey}
\affil{$^2$Department of Physics, The George Washington University, Washington, DC 20052, USA}
\affil{$^3$College of Computer, Mathematical, and Natural Sciences, University of Maryland, College Park, MD 20742, USA}
\email{edasonbas@yahoo.com}


\begin{abstract}
We present the spectral analysis of an 87~ks \emph{XMM-Newton} observation of Draco, a nearby dwarf spheroidal galaxy. Of the approximately 35 robust X-ray source detections, we focus our attention on the brightest of these sources, for which we report X-ray and multiwavelength  parameters. While most of the sources exhibit properties consistent with AGN, few of them possess characteristics of LMXBs and CVs. Our analysis puts constraints on population of X-ray sources with $L_X>3\times10^{33}$~erg~s$^{-1}$ in Draco suggesting that there are no actively accreting BH and NS binaries. However, we find 4 sources that could be LMXBs/CVs in quiescent state associated with Draco. We also place constraints on the central black hole luminosity and on a  dark matter decay signal  around 3.5~keV. 

\end{abstract}

\keywords{galaxies: individual (Draco dwarf) --- galaxies: dwarf --- X-ray: binaries --- novae, cataclysmic variables --- stars: carbon ---  dark matter}

\section{Introduction}

The large spirals of the Local Group, the Milky Way and Andromeda (M31), are known to have a significant number of faint satellite galaxies; some of the faintest discovered most recently by the Sloan Digital Sky Survey (SDSS; e.g., \citealt{2002ApJ...569..245N,2002AJ....123..848W}). The dwarf spheroidals (dSphs) are by far the most numerous class among these satellites and typically the least luminous. In general, these objects are gas-poor, spatially diffuse, and exhibit significantly elevated mass-to-light ratios ($M/L\sim few\times100$; \citealt{1998ARA&A..36..435M, 2008MNRAS..391..942B}). The latter implies that the dSphs occupy an important position in the mass spectrum representing the smallest known structures where dark matter dominates. The dSphs have  low mean metallicity, and thus their study can provide a probe of stellar evolution at these low metallicities. In addition, these objects are of interest because they may have abundance ratios that are different from those of the Milky Way (\citealt{2011ApJ...727...78K, 2011ApJ...727...79K}).

\begin{figure*}
\begin{center}
\includegraphics[scale=0.35]{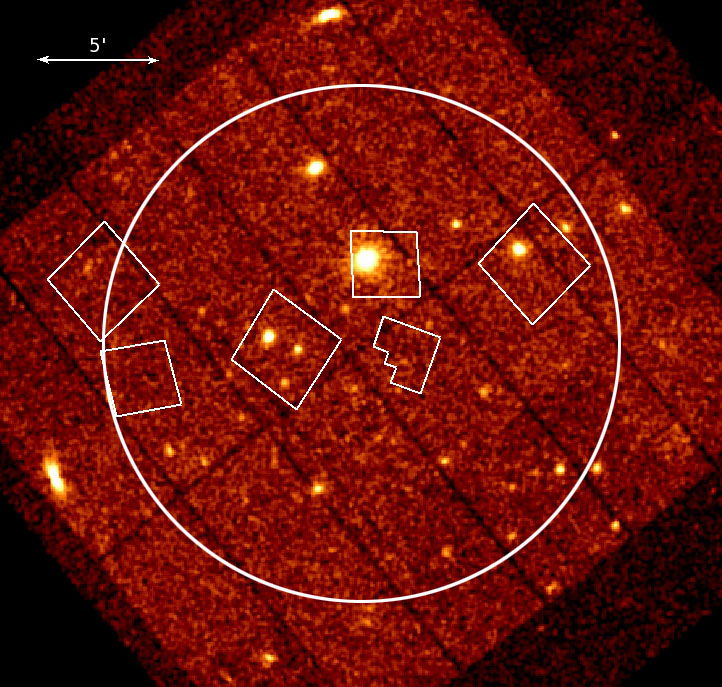}
\includegraphics[scale=0.35]{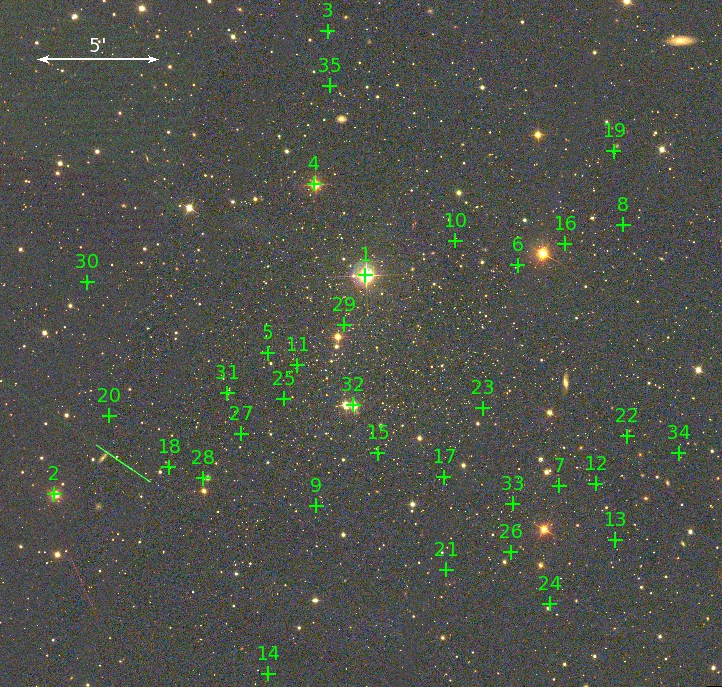}
\caption{{\it Left:} EPIC image of Draco smoothed with gaussian, $r=2''$ kernel. The \emph{HST} pointings are shown in white. With a tidal radius of $\sim40'$, Draco extends beyond the XMM field of view \citep{2001AJ....122.2538O}. The half-light radius of $\sim$$10'$ \citep{2007ApJ...663..948G} is shown with white circle. {\sl Right:} Color-coded (Blue: $u$, Green: $g$, and Red: $r$) SDSS image of Draco. The X-ray sources are shown with green crosses, and numbered according to Table~3. Both images show the same region of the sky (north is up, east is to the left).}
\end{center}
\end{figure*}

\begin{figure*}
\begin{center}
\includegraphics[scale=0.135]{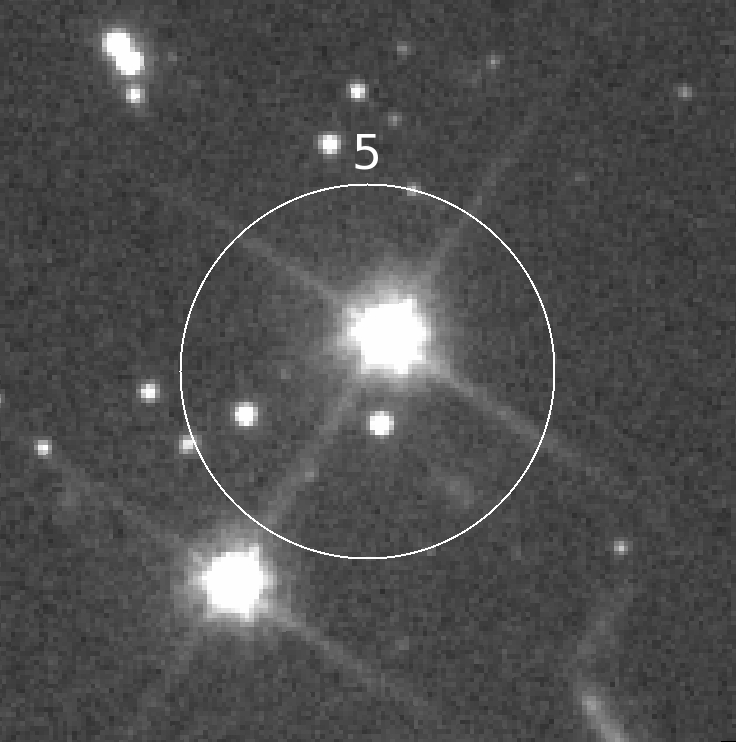}
\includegraphics[scale=0.135]{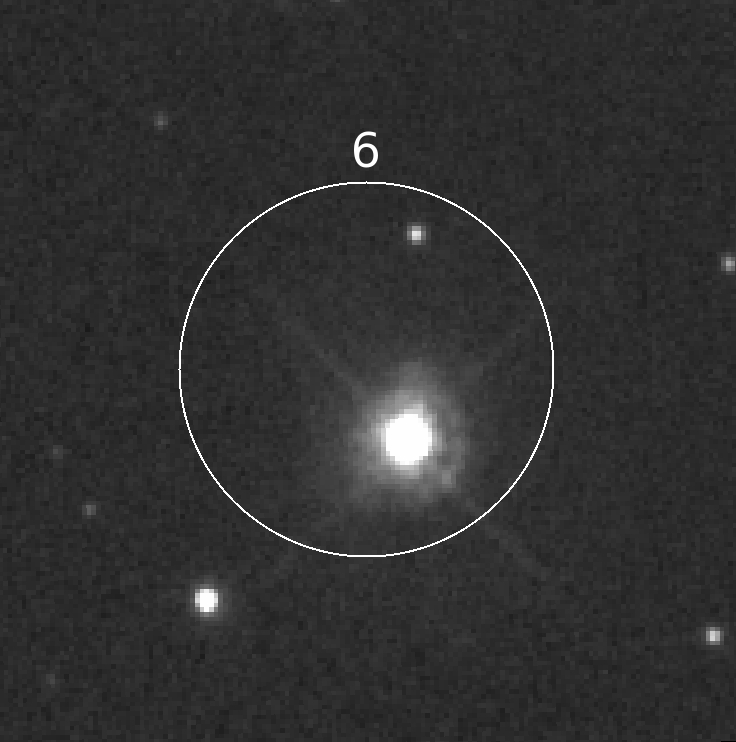}
\includegraphics[scale=0.135]{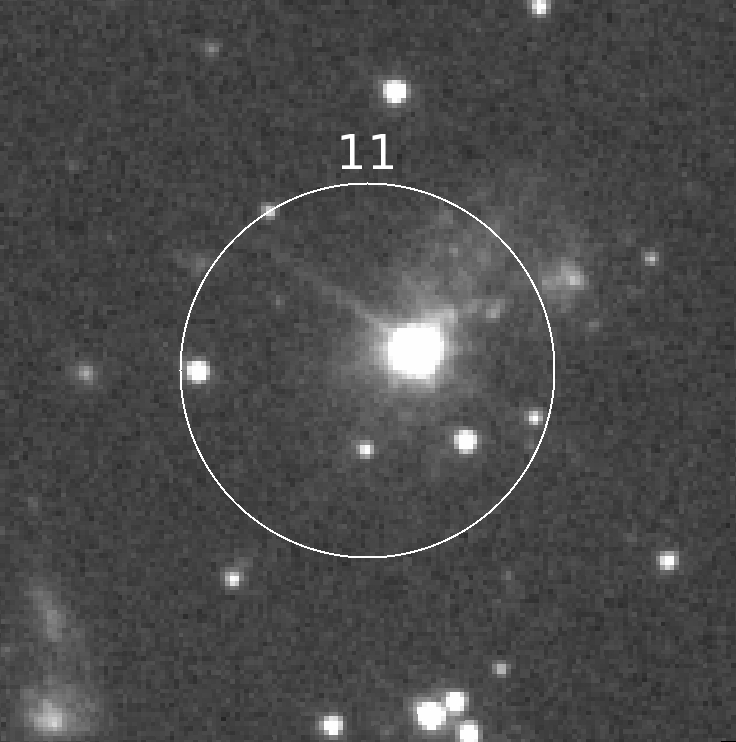}
\includegraphics[scale=0.135]{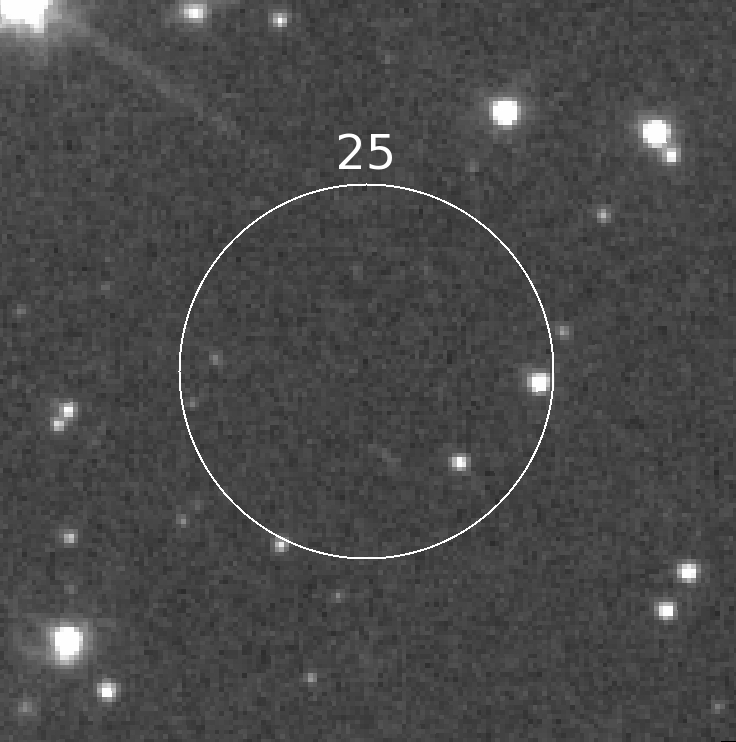}
\includegraphics[scale=0.135]{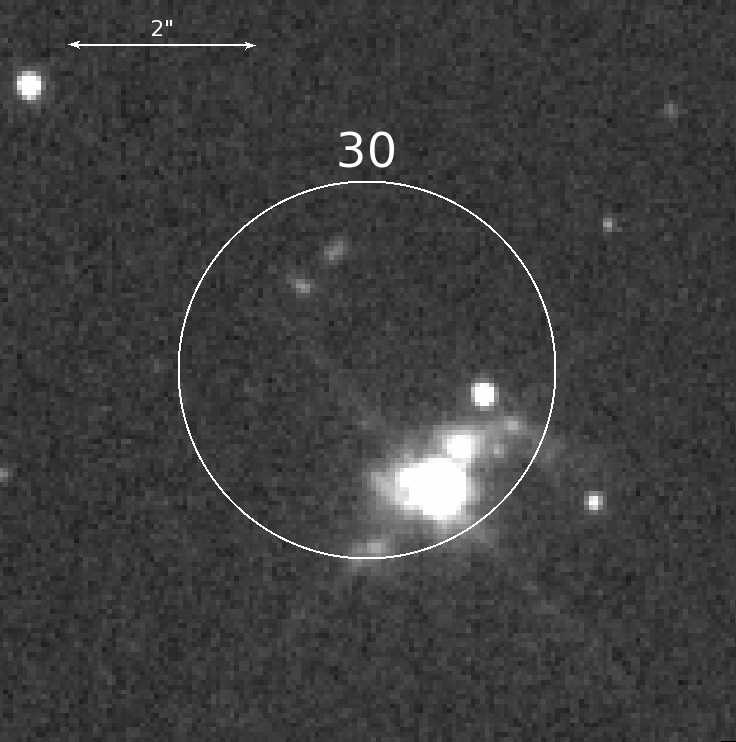}
\caption{\emph{HST} cutouts ($8''\times8''$) centered on the X-ray sources observed by \emph{HST}. Each circle has a $2''$ radius, which represents the typical positional uncertainty for these {\sl XMM-Newton} sources. Note that for source 1, a very bright foreground star is also seen in the \emph{HST} images, and is not shown because the image would be flooded with the light from the star.}
\end{center}
\end{figure*}


A recent study on the Sculptor dwarf galaxy reported the detection of five X-ray binaries with L$_{X}> 6\times10^{33}$ erg~s$^{-1}$ \citep{2005MNRAS.364L..61M}. The number seems to be surprisingly high given the expectations based purely on scaling and integrated properties such as central density, core radius, and velocity dispersion. However, this number appears to be in agreement with predictions of the model proposed by Piro et al.\ (2002), where the discovered X-ray sources are potentially low duty cycle transients. By assuming typical velocities for LMXBs, Dehnen \& King  (2006) used the X-ray data to place a mass limit of $>10^9$~M$_{\odot}$ on the amount of dark matter (DM) in Sculptor required to retain the LMXBs. In addition, the authors note that there should be an extended halo of qLMXBs which may be observable. Simulation studies, involving tidal stripping, (Read et al. 2006b), suggest comparable properties for Draco.\\ 
\\
Located at $\sim80$~kpc, Draco is a faint, metal-poor ($[{\rm Fe/H}] =-1.8\pm0.2$~dex) dSph with an old ($8-10$~Gyr) stellar population \citep{2007MNRAS.375..831S}. \citet{2001AJ....122.2538O} show that Draco's profile is well fit with a King model \citep{1962AJ.....67..471K} with a core radius of $7\farcm7$, and a tidal radius of $40\farcm1$. Under the assumption of virial equilibrium, the high stellar velocity dispersion implies an extremely high mass-to-light ratio (M/L) of about $146\pm42$. In order to probe whether the main drivers of scaling in dSphs are the internal collisions and encounters, which strongly depend on the (core) density of these systems, a population census of LMXBs, qLMXBs and cataclysmic variables (CVs) is highly desirable. This was one of the main goals of our \emph{XMM-Newton} observation whose results we report in this paper.

The paper is organized as follows. Section 2 reports the details of observations, the data reduction and the source selections criteria. In Section 3, we outline our source classification methodology. In Section 4, the spectral analysis is described for the interesting sources, limits on the X-ray emission from the central black hole are presented, and the diffuse emission spectrum near 3.5~keV is presented. In section 5, we summarize our results.

\section{Observations and data analysis}

\subsection{XMM-Newton}

The 87~ks image of Draco (Fig. 1; $RA=17^h20^m12.4^s$ and $DEC=+57^\circ54'55\farcs3$) was obtained by the European Space Agency's (ESA) \emph{X-ray Multi-Mirror Mission -- Newton} (\emph{XMM-Newton}) between August 4 - 28, 2009 (PI: K.\ Dhuga; see Table~1 for details).

\begin{figure}
\begin{center}
\includegraphics[scale=0.53, trim=60 130 0 30]{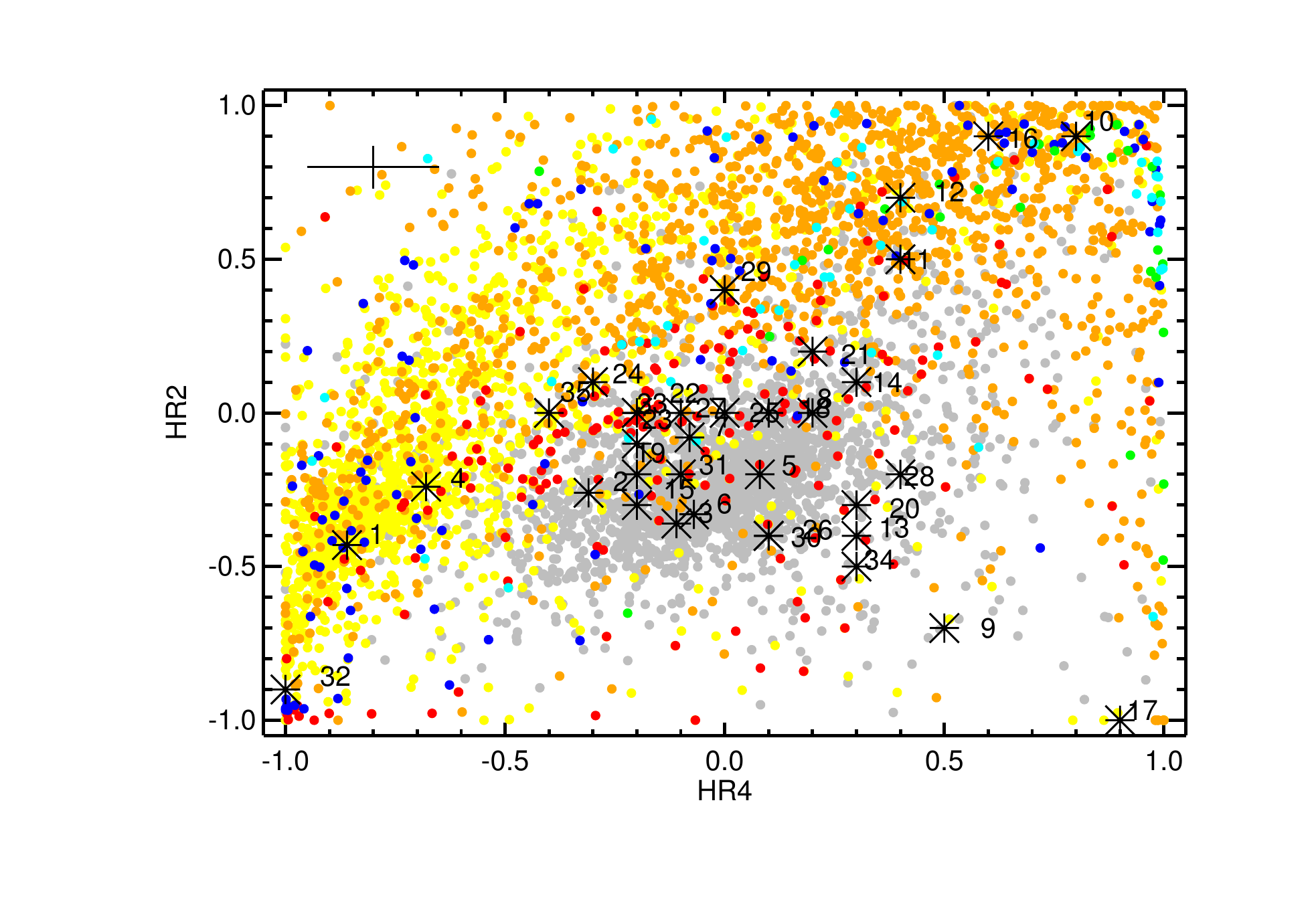}
\vspace{1.2cm}
\caption{HR diagram where sources from the Draco field are shown by asterisks with numbers (correspond to those in Table~3). The population of AGNs from our training dataset are shown in grey, Stars in yellow, YSOs in orange, LMXBs in cyan, HMXBs in green, CVs in red, and pulsars in blue. The average uncertainty of the sources in Table~3 is shown in the top left corner.}
\end{center}
\end{figure}

We used the 3XMM-DR5\footnote{http://xmmssc.irap.omp.eu/Catalogue/3XMM-DR5/3XMM-DR5\_Catalogue\_User\_Guide.html\#DocHistory} (3XMM) catalog \citep{2015arXiv150407051R} for the automated multiwavelength (MW) classification described in Section~3.1. The parameters obtained from the 3XMM catalog are the source coordinates and X-ray fluxes in 4 different energy bands.

For several of the brighter X-ray sources and for the diffuse background emission (see Section~4) we performed spectral analyses and extracted spectra from the EPIC data. The observation data files (ODFs) were processed using standard procedures of the \emph{XMM-Newton} Science Analysis System (XMM-SAS version 13.0.1). The standard SAS tool \texttt{edetect-chain} was used to perform source detections. We extracted spectra using circular ($r=17''$) regions centered on the X-ray sources, for each observation. Background subtraction was performed using source-free apertures located nearby. A standard event screening described in the \texttt{multixmmselect} manual\footnote{http://xmm.esac.esa.int/sas/current/documentation/threads/ multiexposures\_thread.shtml} was applied. The spectral analysis of the X-ray data was performed using XSPEC version 12.7. The pn and MOS spectral energy channels were grouped to have at least 10 counts per bin. Each spectrum was fitted in the 0.2$-$10 keV energy range. 

\begin{deluxetable}{crrc}
\tablecaption{X-ray observations of Draco}
\tablehead{ \colhead{Date} & \colhead{ObsId} & \colhead{Observatory} & \colhead{Exp\tablenotemark{a}} }
\startdata
2009 Aug 04 & 0603190101 & \emph{XMM-Newton} & 17 \\
2009 Aug 06 & 0603190201 & \emph{XMM-Newton} & 18 \\
2009 Aug 08 & 0603190301 & \emph{XMM-Newton} & 16 \\
2009 Aug 20 & 0603190401 & \emph{XMM-Newton} & 18\\
2009 Aug 28 & 0603190501 & \emph{XMM-Newton} & 18
\enddata 
\label{table:nonlin} 
\tablenotetext{a}{Exposure in ks.}
\end{deluxetable}

\begin{deluxetable}{cccccc}
\tablecaption{\emph{HST} observations of Draco}
\tablehead{
\colhead{Date} & \colhead{Instrument} &  \colhead{Filter} & \colhead{ObsId} &  \colhead{Exp.\tablenotemark{a}} & \colhead{Res.\tablenotemark{b}}}
\startdata
2001 Aug 18 & WFPC2 & F555W & 9043 & 680 & 0.1 \\
2004 Oct 31 & ACS\tablenotemark{c} & F606W & 10229  & 8170 & 0.5 \\
2004 Oct 30 & ACS\tablenotemark{c} & F555W & 10229  & 8170 & 0.5 \\
2004 Oct 29 & ACS\tablenotemark{c} & F606W & 10812  & 7200 & 0.5 \\
2012 Oct 26 & WFC3\tablenotemark{d} & F606W & 12966 & 3152 & 0.46 \\
2013 Oct 14 & WFC3\tablenotemark{d} & F606W & 12966 & 3076 & 0.46
\enddata
\tablenotetext{a}{Exposure time in seconds.}
\tablenotetext{b}{Pixel scale in $''$/pixel.}
\tablenotetext{c}{WFC camera.}
\tablenotetext{d}{UVIS camera.}
\end{deluxetable}

\subsection{Hubble Space Telescope}

The \emph{Hubble Space Telescope} (\emph{HST}), partly observed Draco with the WFPC2, ACS/WFC or WFC3/UVIS cameras (Table~2 lists the images used for this study). 
Figure 1 shows the \emph{HST} pointings, in white, overlaid on top of the EPIC image. Figure~2 shows $8''\times8''$ cutouts centered on X-ray sources observed by {\sl HST}. We use the high-resolution \emph{HST} images to distinguish between AGNs and stars.

\section{Multiwavelength Analysis}

\subsection{Classification}

Traditional multiwavelength (MW) classification includes heuristic examination of the MW source properties, such as X-ray hardness ratios, X-ray/optical/NIR flux ratios, color-color, and color-magnitude diagrams in
the optical and IR such as those shown in Figures 3, 4 and 5 respectively. These diagrams allow one to devise  crude criteria to discriminate between the sources of different nature (e.g., \citealt{2006ApJS..163..344K,2010ApJ...725..931M,2012ApJ...745...99K,2012ApJ...756...27L}). While the MW analysis is often the only reliable way to unveil the nature of unknown X-ray sources, traditional classification, based on drawing dividing lines in the MW diagrams is a very laborious process lacking any quantitative classification confidence criterium. 

There are over a hundred X-ray sources in the 3XMM catalog\footnote{http://xmmssc-www.star.le.ac.uk/Catalogue/3XMM-DR4/ UserGuide\_xmmcat.html} in the Draco field. We visually inspected all sources, and after excluding source duplications (in 5 separate observations of Draco) and spurious detections (e.g., edge of the chip, bad CCD columns) we selected 35 bright X-ray sources with $S/N>6$ for further investigation. Table~3 lists the source parameters: X-ray flux $F_{0.5-12}$ in $0.5-12$~keV range and two hardness ratios HR2 and HR4 defined as $HR2 = (F_{(1.2-2} - F_{0.2-1.2})/(F_{1.2-2} + F_{0.2-1.2})$ and $HR4 = (F_{2-7} - F_{0.5-2})/(F_{2-7} + F_{0.5-2})$, where $F_{x-y}$ are the observed fluxes in the respective energy bands. Figure~3 shows the hardness ratio diagram for the sources in Draco compared to different classes of literature verified sources used for automated classification (see A1 for details). The two chosen hardness ratios allow for a some separation between AGNs and the rest of the sources, although there is an expected overlap between the AGNs, CVs, and pulsars.

Similarly, we used color-color diagrams to compare the sources in Draco to known sources of different classes (AGN, CV, pulsars etc.). The left panel of Figure 4 shows a good degree of separation between the stars and AGN in terms of  IR to X-ray flux ratio\footnote{Note that in Figure 4 (left panel) we plot the spectral flux ratios which differ from the flux ratios by the factor equal to the inverse ratio of the corresponding mid-band frequencies.}, which is often used to separate the two classes.  The color-color diagram in the right panel of Figure 4 also shows clear separation between the AGN and stars but it also allows to see noticeable separation between the AGN and CVs. We also utilize color-magnitude diagrams (CMDs; see Figure~5) and visually inspect the \emph{HST} and SDSS images. The former allows us to estimate the age of stellar sources, which in turn helps to determine whether a given star is part of Draco. If a star, whose absolute magnitude is calculated for the Draco distance, happens to land onto a $<1$~Gyr isochrone, it is unlikely to be part of Draco, because the stellar population of the dSph is at least several billion years old \citep{2007MNRAS.375..831S}. AGNs have a spectral energy distribution (SED) that is different from that of stars, and, therefore, on different CMDs, AGNs ``jump'' between different isochrones, or do not land on one at all.

\begin{figure*}
\begin{center}
\includegraphics[scale=0.5, trim=60 50 0 30]{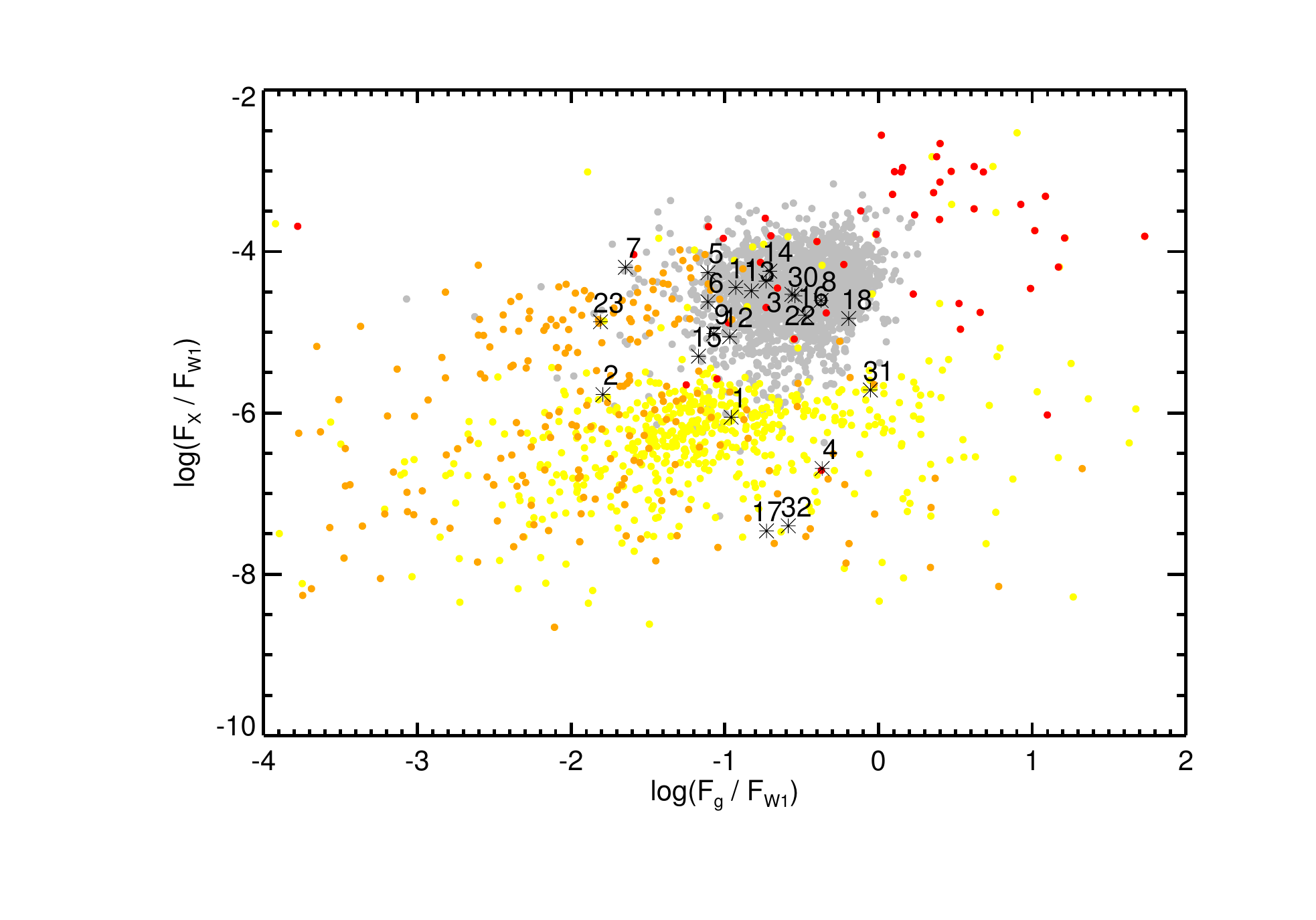}
\includegraphics[scale=0.5, trim=60 50 0 30]{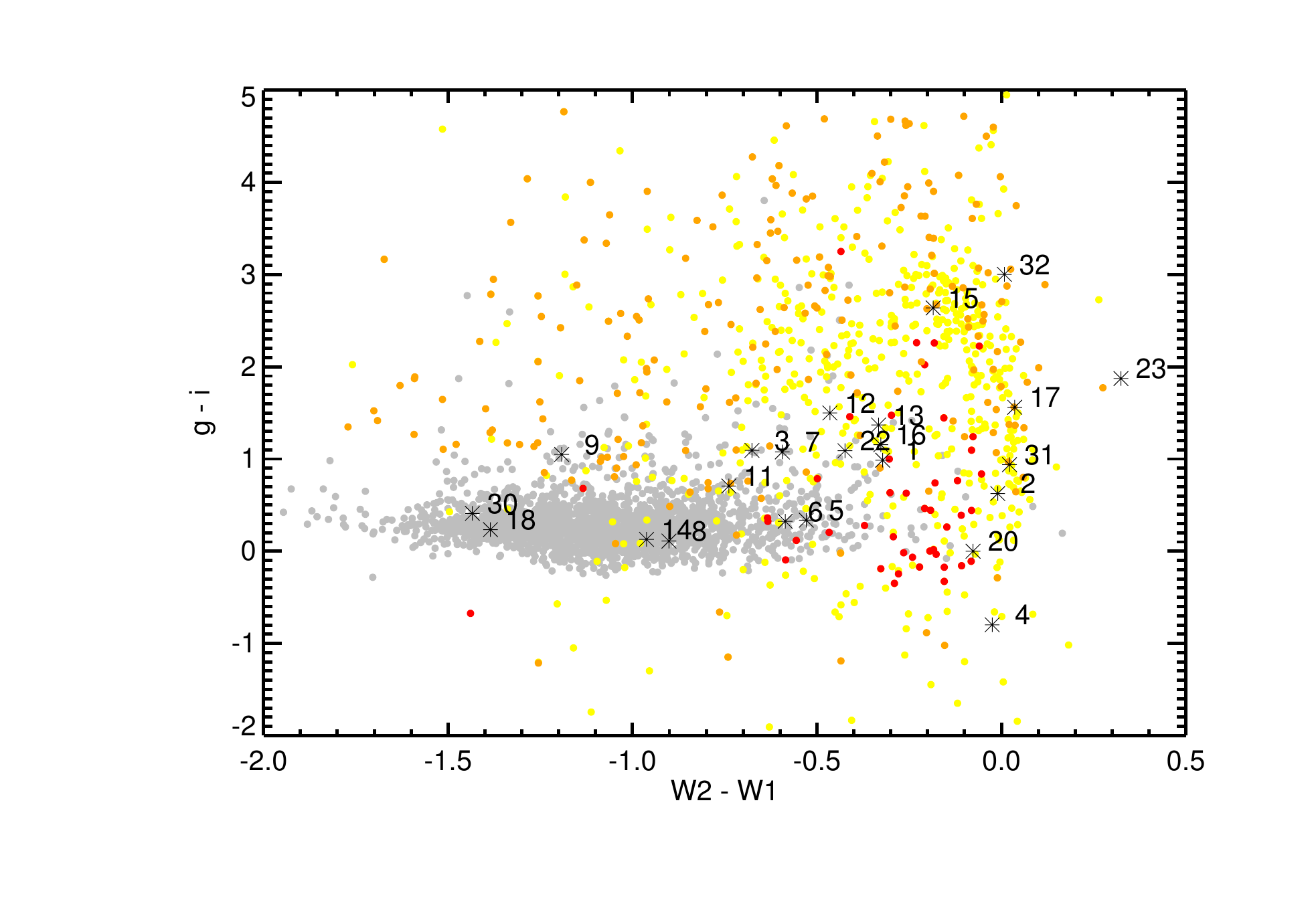}
\caption{The flux ratio and color-color diagrams traditionally used for source classification. These also represent of the many possible slices in MW parameter space used by the automated algorithm to create the decision tree (see Appendix). {\bf Left}:  X-ray-to-IR spectral flux (at mid-band frequency) ratio, $F_{\rm 0.5-8 keV} /F_{W 1}$, vs.\  optical-to-IR spectral flux ratio, $F_g/F_{W1}$, is shown in logarithmic scale. {\bf Right:} Optical color, $g-i$  vs.\ IR color, $W2-W1$. On both panels the population of AGNs from the  training dataset (see Appendix) is shown in grey, YSOs in orange, stars in yellow, and CVs in red. The sources from the Draco field (only those with the optical and NIR counterparts)  are shown as black asterisks and numbered according to Table~3.}
\end{center}
\end{figure*}

\begin{deluxetable*}{llllllllllll}
\tablecaption{Properties and automated results for classification X-ray sources in the Draco field}
\tablehead{\colhead{\#} & \colhead{RA} & \colhead{Dec} & \colhead{$\delta r$\tablenotemark{a}} & \colhead{Flux\tablenotemark{b}} & \colhead{HR2\tablenotemark{c}} & \colhead{HR4\tablenotemark{c}} & \colhead{S/N\tablenotemark{d}} & \colhead{$g$\tablenotemark{e}} & \colhead{$i$\tablenotemark{e}} & \colhead{W1\tablenotemark{f}} &  \colhead{Class\tablenotemark{g} (Prob.)}}
\startdata
1&260.0917&57.9740&0.40&86(2)&-0.43(1)&-0.86(1)&115.4&12.80&11.81&7.73&  GM Dra\tablenotemark{h} \\
2&260.4930&57.8228&0.56&157(16)&-0.26(2)&-0.31(6)&51.3&11.21&10.10&8.25&   Star/CV ? \\
3&260.1406&58.1417&0.58&74(8)&-0.36(2)&-0.11(6)&48.4&17.34&16.24&12.83&   AGN (84\%)\\
4&260.1575&58.0366&0.49&21(2)&-0.24(3)&-0.68(6)&43.4&11.39&12.19&7.80&   Star\tablenotemark{i} (99\%)\\
5&260.2180&57.9202&0.47&18(2)&-0.20(4)&0.08(7)&31.8&20.29&19.96&14.84&   AGN (99\%)\\
6&259.8937&57.9806&0.47&15(3)&-0.33(4)&-0.1(1)&26.1&19.34&19.02&13.90&   AGN (99\%)\\
7 &259.8409&57.8290&0.52&5(2)&-0.08(8)&-0.1(2)&16.1&22.76&21.69&15.98& ? \\
8&259.7567&58.0081&0.67&6(2)&0.0(1)&0.2(2)&14.4&18.79&18.68&15.18&   AGN (97\%)\\
9 &260.1548&57.8155&0.79&12(3)&-0.73(6)&0.5(1)&14.3&19.32&18.28&13.97&   AGN\tablenotemark{j} (99\%)\\
10 &259.9750&57.9977&0.71&1(2)\tablenotemark{m}&0.91(1)&0.78(4)&14.2&22.97&21.88&&   ? \\
11 &260.1798&57.9119&0.69&5(1)&0.5(1)&0.4(1)&14&20.93&20.22&15.93& CV/AGN \\
12 &259.7936&57.8301&0.75&10(2)&0.69(9)&0.44(9)&13.4&18.96&17.46&13.87&   AGN\tablenotemark{j} (88\%)\\
13 &259.7689&57.7914&0.81&5(2)&-0.4(1)&0.3(2)&12.5&20.49&19.12&15.75&   AGN (99\%)\\
14&260.2161&57.6997&0.86&15(5)&0.1(1)&0.3(2)&12.5&19.85&19.72&15.41&   AGN (99\%)\\
15&260.0755&57.8516&1.65&2.6(8)&-0.3(1)&-0.2(2)&12.4&19.48&16.85&13.88&   AGN (84\%)\\
16 &259.8326&57.9952&0.85&9(2)&0.9(1)&0.61(8)&12&18.80&17.68&14.97&   AGN ?\\
17&259.9902&57.8351&0.86&0.1(4)\tablenotemark{m}&-1.0($^{+0.3}_{-0}$)\tablenotemark{n}&0.9($^{+0.1}_{-0.9}$)\tablenotemark{n}&10.6&17.75&16.19&13.26&   C1\tablenotemark{k} \\
18&260.3452&57.8417&1.02&5(2)&-0.0(1)&0.1(2)&10.4&18.05&17.81&14.89&   AGN (84\%)\\
19 &259.7686&58.0586&0.86&6(3)&-0.2(1)&-0.2(3)&10&  &  &&   ? \\
20&260.4225&57.8769&1.11&9(3)&-0.3(1)&0.3(2)&10&  &  &15.75& AGN? \\
21&259.9869&57.7715&1.17&5(2)&0.2(1)&0.2(2)&9&  &  &&   ? \\
22&259.7530&57.8628&0.99&2(1)&0.0(1)&-0.2(3)&8.8&20.27&19.18&16.19&   AGN (99\%)\\
23 &259.9394&57.8825&1.05&0.9(5)&-0.1(1)&-0.2(3)&8.6&23.20&21.33&16.01&   ? \\
24&259.8536&57.7479&1.14&2(1)&0.1(2)&-0.3(4)&8.3&  &  &&   ? \\
25&260.1971&57.8889&1.17&1.3(9)&-0.0(2)&0.0(4)&7.9&  &  && ? \\
26&259.9040&57.7832&1.16&1.3(9)&-0.4(2)&0.1(3)&7.9&22.84&21.35&& AGN? \\
27&260.2515&57.8649&1.05&1(1)&-0.0(4)&-0.1(2)&7.6&21.45&21.53&&  AGN? \\
28&260.3006&57.8342&1.06&3(2)&-0.2(2)&0.4(2)&7.6&22.99&23.18&&  ?\\
29&260.1186&57.9400&1.20&0.6(5)&0.4(2)&0.0(3)&7.3&  &  && ? \\
30&260.4516&57.9685&2.00&2(3)\tablenotemark{m}&-0.4(2)&0.1(5)&7.2&20.22&19.81&16.19& AGN (99\%)\tablenotemark{l}\\
31&260.2698&57.8927&1.21&2(2)&-0.2(2)&-0.1(4)&7.2&15.97&15.03&13.17&   Star (99\%)\\
32&260.1072&57.8848&1.16&0.3(3)&-0.9(1)&-1.0($^{+0.5}_{-0}$)\tablenotemark{n}&7.1&14.30&11.30&10.16&  Star (95\%) \\
33&259.9014&57.8166&1.18&0.9(6)&-0.0(2)&-0.2(3)&6.8&  &  &&   ? \\
34&259.6866&57.8509&6.27&6(6)&-0.5(2)&0.3(4)&6.2&  &  && ? \\
35&260.1380&58.1042&1.69&1(1)&0.0(2)&-0.4(6)&6.1&  &  &&   ?
\enddata 
\tablecomments{$^a 1\sigma$ positional uncertainty in arcseconds (from 3XMM-DR5 catalog). $^b$Observed X-ray fluxes in the $0.5-12$~keV range in units of $10^{-14}$  erg~s$^{-1}$~cm$^{-2}$. The value in brackets is the measurement uncertainty in the last digit. 
$^c$Hardness ratio calculated as $HR2 = (F_{(1.2-2} - F_{0.2-1.2})/(F_{1.2-2} + F_{0.2-1.2})$ and $HR4 = (F_{2-7} - F_{0.5-2})/(F_{2-7} + F_{0.5-2})$, where $F_{X-Y}$ is the observed flux in the  $X-Y$ keV energy band. The value in brackets is the measurement uncertainty in the last digit. $^d$Signal-to-noise ratio. $^eg$ and $i$ magnitudes front the USDSS DR9 catalog. $^f$W1 magnitudes front the WISE catalog. $^g$Classification confidence (see text). $^h$GM Dra, variable star in Tycho 2 catalog \citep{{2000A&A...355L..27H}}. $^i$A cool K9 type star with X-ray luminosity of $3.7\times10^{29}$\,ergs\,s$^{-1}$ (at a distance of 125~pc; \citealt{2006ApJ...638.1004A}). $^j$Visually resolved in the SDSS images.$^k$Symbiotic star in Draco, see \citet{2000A&AS..146..407B}. $^l$Source \#30 also shows complex morphology (likely, superposition of star and galaxy as can be seen in Figure~2. $^m$Three sources have the uncertainties of fluxes in the individual bands (1$-$2 keV, 2$-$4.5 keV, etc.) larger than the fluxes themselves (according to 3XMM-DR5), which results in the large  total flux uncertainty. 
$^n$ In these cases the uncertainty is strongly asymmetric, because the value of HR hits the boundary, and therefore two uncertainties are provided.}
\end{deluxetable*}

\begin{figure}
\begin{center}
\includegraphics[scale=0.45,trim=0 40 0 30]{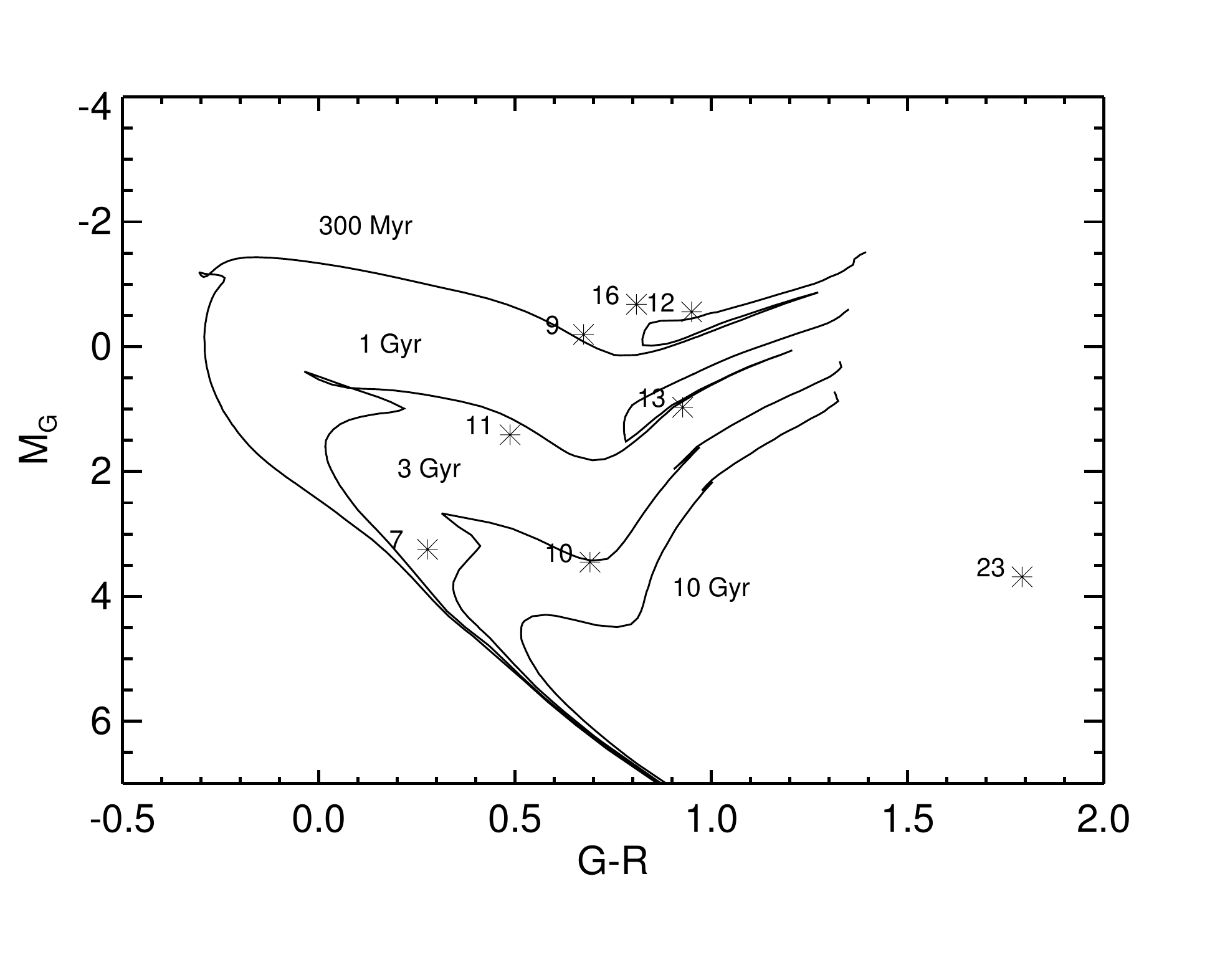}
\caption{Color-magnitude diagram showing absolute magnitude $M_G$ vs. color G-R. Isochrones are plotted at 300 Myr, 1 Gyr, 3Gyr, and 10 Gyr. Sources from Table~3 are shown with asterisks.}
\end{center}
\end{figure}

Since we are interested in objects that belong to Draco we attempt to eliminate the Galactic stars. One helpful constraint is the optical brightness of these objects at the known distance to Draco. The distance modulus of Draco (80\,kpc) is $\sim19.5$. Figure~5 clearly shows that we do not expect to see any stars brighter than $M_g\approx-1$ (a very conservative estimate, which translates to apparent magnitude $m_g\approx18$), assuming a few Gyr as the age of Draco. This means that stars brighter than $m_g\approx 18$ likely belong to our Galaxy rather than Draco. (This criteria does not apply to QSOs/AGNs, which can be brighter.) Sources \#1, \#2, \#4, \#11, \#31 and \#33 appear to coincide with stars associated with our Galaxy. Inspection of the \emph{HST} images confirms that source \#1 also contains a very bright foreground star. 

In addition, we utilize the superb angular resolution of \emph{HST} to identify AGNs (we also check the lower-resolution  SDSS images when {\sl HST} image is not available). Unfortunately, only a very limited (often single-band) \emph{HST} coverage of Draco currently exists (see Figure~1, left panel). Nonetheless, high resolution \emph{HST} images allow us to eliminate several foreground and background sources to Draco (see Table~3).

Sources \#5 and \#6 appear to be point-like in the {\sl HST} images, but given their optical magnitudes (of 19-20) we cannot exclude the possibility of QSOs (which could also appear point-like). Based on the SDSS images and photometry (see  Figure~4, right panel), we also conclude that \#2 and \#32 are likely (bright) foreground stars. The SDSS images suggest that \#12 and \#9 are AGNs (in agreement with Figure~4).

We also attempted to verify the classification using the SIMBAD catalog\footnote{http://simbad.u-strasbg.fr/simbad/}. However, we found that the SIMBAD classifications are incorrect for several faint sources. For example, we find AGNs (sources \#9 and \#12, resolved with \emph{HST}) to be classified as stars. Also, a few stars are referred to as QSOs/AGNs in SIMBAD. Therefore, we do not rely on the SIMBAD for classifications of the X-ray sources in Draco.

As an aid to help with the classification, we have developed an automated machine learning classification pipeline (MUWCLASS; \citealt{2014AAS...22325513B}) that we subsequently applied to the X-ray sources in Draco. The pipeline uses a learning decision tree algorithm, to perform the X-ray source classification (see A1 for details). The automated procedure produced 20 classifications with confidence $>70\%$, including 14 AGN, 5 stars, and 1 CV (see Table~3). The classifications with lower confidence are indicated by a question mark. Note that the calculated confidences for each type do not include the uncertainties associated with the X-ray flux determination which are substantial for faint X-ray sources and also affect {\sl HR}s. They also do not include a possibility of assigning false MW counterparts to the X-ray sources (confusion), which, however, is expected to be small (see Section~3.2). For a number of interesting X-ray sources, which could belong to Draco or be XRBs or compact objects in our Galaxy, we have carried out a more detailed analysis (see Section~4.1) by extracting the X-ray spectra from the original data and performing spectral fits (not used in the automated classifications). The locations of the sources in the diagrams shown in Figures 3$,$4 and 5  are mostly consistent with the automated classifications or, whenever this is not the case, the reasons for discrepancy are discussed in Section~4.1 on source-by-source basis.

After we submitted this paper, we became aware of two other papers (Manni et al.\ 2015 and Saeedi et al.\ 2015) discussing X-ray source classification in Draco and  submitted  roughly at the same time as our paper. The Saeedi et al.\  (2015) paper presents a more comprehensive analysis than Manni et al.\ (2015) and, bearing in mind differences in class definitions, nearly all our classifications agree with Saeedi et al.\ (2015),  except for source 15 for which our automated  tool preferred AGN (84\%) classification, while Saeedi et al.\ (2015) suggest a binary star (not XRB) classification (but  uncertain). We also note that the machine-learning method used in our paper  is different from those used by  Saeedi et al.\ (2015) and Manni et al.\ (2015)  who relied on previously established heuristic prescriptions from multiple publications. However, the very good agreement between our and Saeedi et al.\ (2015) classifications shows that both methods worked well for this field and  similar results were obtained with different classification  approaches, adding credibility to both independent analyses.

\subsection{Cross-correlation and chance superposition}

We cross-correlated all X-ray sources listed in Table~3 with the USNO-B1, SDSS, 2MASS, and WISE catalogs.  We find a single optical/NIR/IR counterpart within the $2''$ search radius\footnote{This happens to be an upper bound on 1$\sigma$ position uncertainties for 34 out of 35 X-ray sources in the Draco field based on the 3XMM-DR5 positional parameters (see Table~1).} for all but one source\footnote{This is source \#34 which is both very faint and strongly off-axis.} X-ray source. The only sources with multiple counterparts within the search radius are source \#5, which has two USNO-B1 counterparts, and source \#32, which also has two counterparts in SDSS. For those two sources we only consider the brightest (which also happens to be the closest) counterpart to be the ``true''  counterpart below. We have verified that all optical/NIR/IR counterparts are within $\sim$0\farcs5 of one another. We  see no systematic offset between the X-ray and 2MASS sources and the RMS offset between the X-ray and 2MASS/SDSS positions for the 35 sources listed in Table~3 is $\approx1''$, i.e.\ well within the distribution of uncertainties for the X-ray source positions (see Table~3). We note that while the \emph{HST} images reveal multiple counterparts for some of the sources, the surveys are not as deep, and, therefore, pick up only the brightest objects in the $2''$ search radius around each source. Unfortunately only 5 X-ray sources are located within the \emph{HST} fields and even for them multiband {\sl HST} photometry is lacking. We cannot exclude that some of the brighter X-ray sources are associated with the much fainter optical sources that can only be seen in the {\sl HST} images. However, these would have to be sources of a very rare type rare types (e.g., solitary neutron stars or quiescent BH binaries\footnote{such a Swift J1357.2$-$0933} in Draco or extended  Galactic halo)  or yet unknown classes. As for the fainter X-ray sources, we can hardly say much without deep {\sl XMM-Newton} and {\sl HST} observations. 

For each survey we calculate the chance superposition using three different techniques. First, based on the average optical/NIR/IR source densities in the field ($\rho=0.0048$, 0.01, 0.0004, and 0.001 stars/arcsec$^2$ in the USNO-B1, SDSS, 2MASS, and WISE surveys, respectively) we calculate the probability of finding finding zero field sources in the circle of radius $r=2''$, $P = \exp(-\rho\pi r^2)$. This leads to chance coincidence probabilities of  $1-P=6\%$, 12\%, 0.5\%, and  1\% (for USNO-B1, SDSS, 2MASS, and WISE surveys, respectively). For the second method we run 1,000 simulations where we randomly populate with 35 synthetic X-ray sources the optical/NIR/IR field of Draco (similar to e.g., \citealt{2012ApJ...758...99R}). We then count the average number of optical/NIR/IR counterparts to the synthetic X-ray sources within the $r=2''$ radius. This method results in 0.9, 1.9, 0.08, and 0.27 (for USNO-B1, SDSS, 2MASS, and WISE surveys, respectively) spurious detections of the 35 sources. The third method we employed was to offset each X-ray sources in random direction (within $100''$ of its original position). This led to 1, 1.8, 0.1, and 0.4 (for USNO-B1, SDSS, 2MASS, and WISE surveys, respectively) spurious detections in 1,000 simulations. Therefore, all three methods give very similar results suggesting only 1-2 spurious detections, compared to the 26 X-ray sources for which we find optical/NIR/IR counterparts (see Table~3).

\section{Results and Discussion}

Our goals are to identify and/or set constraints on the numbers of XRBs that may belong to Draco, set a limit on the emission from a possible central intermediate mass BH, and look for a possible DM decay line in the diffuse X-ray emission.

\subsection{X-ray sources}

Below, we discuss individually several sources that do not appear to be AGNs or foreground Galactic stars based on the color-color and color-magnitude diagrams, HST/SDSS images, and automated classification. For the brightest of these sources we performed spectral fits with  blackbody radiation (\texttt{bbodyrad}), disk blackbody (\texttt{diskbb}) and power law (PL) models modified by interstellar absorption (\texttt{phabs} model in XSPEC). In general these models provide an adequate description of the thermal and non-thermal components of X-ray spectra typically observed in the 0.2$-$10\,keV range for objects such as CVs, isolated neutron stars, and LMXBs. The best-fit parameters for these sources are given in Table~3. 

\begin{deluxetable*}{ccccccccccc}
\tablecaption{Best-fit spectral results for selected point sources in the Draco field.}
\tablehead{ \colhead{Source \#} & \colhead{Model} &\colhead{Normalization\tablenotemark{a}} &\colhead{n$_H\tablenotemark{b} $}&\colhead{\textit{kT}, T$_{in}$\tablenotemark{e}} &\colhead{$\Gamma$} &\colhead{$\chi_{\nu}^2$} & \colhead{$\nu\tablenotemark{f} $}\\
\colhead{} & \colhead{} &\colhead{} &\colhead{(10$^{22}$cm$^{-2}$) } &\colhead{(keV)} &\colhead{} &\colhead{} & \colhead{}
}
\startdata
\#10 &  PL & (1.53$\pm$0.48)$\times10^{-5}$ &	1.35$\pm$0.27	& 	&		1.72$\pm$0.23	& 0.85 & 171\\
	&  diskbb  & (1.92$\pm$1.15)$\times10^{-4}$  & 0.83$\pm$0.16	& 2.09$\pm$0.35		&  & 0.85 & 171\\
	& bbodyrad & (3.35$\pm$0.89)$\times10^{-3}$ & 0.35$\pm$0.14	& 1.11$\pm$0.08& & 	0.88 & 171\\
\#11 & bbodyrad+PL & (5.64$\pm$2.94)$\times10^{-3}$~~\tablenotemark{c}   & $<$ 0.027 & 0.75$\pm$0.11& 1.24$\pm$0.48 & 1.05 &171\\
	&  & (2.20$\pm$1.35)$\times10^{-6}$~~\tablenotemark{d}   & &	&		&	& \\
\#16& PL & (1.85$\pm$0.50)$\times10^{-5}$  & 0.92$\pm$0.17& 	&	1.82$\pm$0.21	& 		1.16  & 60\\
\#19& PL  & (1.01$\pm$0.16)$\times10^{-5}$ &	0.06$\pm$0.04& 	&		2.91$\pm$0.38	& 0.77  & 43\\
	&  bbody & (3.32$\pm$0.28)$\times10^{-7}$  &	$<$ 0.027 & 	0.18$\pm$0.01 & &	1.06  & 43\\
	& bbodyrad  & 2.21$\pm$0.63  & $<$ 0.027& 0.18$\pm$0.001	&		&	1.06 & 43\\
\#23 &  PL & (3.47$\pm$1.98)$\times10^{-6}$ &	0.23$\pm$0.18& 	&		3.6$\pm$1.3	& 0.89	& 75\\
	&  bremss & (5.63$\pm$5.56)$\times10^{-6}$ &	$<$ 0.027 & 0.83$\pm$0.55	&			& 0.88	& 75\\
\#25& diskbb & (2.12$\pm$0.14)$\times10^{-3}$ &	$<$ 0.027 & 0.64$\pm$0.11	&			& 0.83	& 197
\enddata
\tablenotetext{a} {Normalization of the \texttt{bbodyrad} model is R$^{2}_{km}/D_{10}^2$, where R$_{km}$ is the source radius in km, D$_{10}$ is the distance to the source in units of 10\,kpc and the normalization of the {\tt diskbb} model given as $((R_{in}/{\rm km})/(D/10~{\rm kpc}))^2$cos$\theta$, where R$_{in}$ is the is "an apparent" inner disk radius.}
\tablenotetext{b} {Galactic extinction is assumed as 2.7$\times$10$^{20}$ cm$^{-2}$ for each source.}
\tablenotetext{c} {Normalization of the {\tt bbody} component. }
\tablenotetext{d} {Normalization of the PL component.}
\tablenotetext{e} {kT; temperature keV from {\tt bbody} and {\tt bbodyrad} models. $T_{in}$; temperature at inner disk radius (keV) from {\tt diskbb} model.}
\tablenotetext{f} {number of degrees of freedom.}
\label{table:nonlin} 
\end{deluxetable*}

{\bf Source \#10:} The optical-IR properties of this source exhibit characteristics of an old (few Gyr), evolved star. The X-ray spectrum is well fitted with the \texttt{bbodyrad} model with a temperature of of $1.11\pm0.08$ keV. Normalization of the \texttt{bbodyrad} model corresponds to an effective emission region of 0.44 km at 80 kpc. The absorption n$_H=3.5\times10^{21}$\,cm$^{-2}$ exceeds the Galactic $n_{H}=2.69\times10^{20}$\,cm$^{-2}$ in this direction (\emph{Chandra} Colden toolkit\footnote{http://cxc.harvard.edu/toolkit/colden.jsp}). The source \texttt{diskbb} model gives a similar $n_H$ and temperature of $2.1\pm0.3$\,keV. If the source is in Draco, then the normalization of the {\tt diskbb} model, $((R_{in}/km)/(D/10~kpc))^2$cos$\theta$, $\theta$ is the angle of the disk (assumed face-on), corresponds to $r_{in}= 0.11$ km,  which seems to be too small for an accretion disk in an XRB. The X-ray luminosity, $L_X\simeq8\times10^{34}$~erg~s$^{-1}$, is too high for a redback or black widow type binary unless it goes through an outburst episode and switches to an accretion state (see e.g., \citealt{2014ApJ...790...39S}). However, the X-ray lightcurve does not show any significant variability during all five {\sl XMM-Newton} observations. Alternatively, it is  possible that \#10 is a {\em non-accreting} MSP in a Galactic binary with a late type companion at a distance of $\sim1$\,kpc with the X-rays being emitted from the hot polar cap of the NS. However, the large (compared to the Galactic) absorbing column would be puzzling in this scenario. 

{\bf Source \#11:} The X-ray spectrum requires two components, \texttt{bbody} (BB)+PL, as PL or BB models alone do not provide acceptable fit to the data. Both the temperature ($kT=0.7\pm0.1$\,keV) and the photon index ($\Gamma=1.2\pm0.5$) are not uncommon for several object types (e.g., AGN and CV) but even for the  two-component model the data show systematic excess below $0.4$\,keV over the best fit, suggesting that even softer component is needed. Although the luminosity of  $4.5\times10^{32}$~erg~s$^{-1}$ at $d=80$~kpc is typical of a redback or black widow system, the spectrum is unusually soft for these kind of binaries. The absorbing column is uncertain and low, consistent with the Galactic value. The \emph{HST} observations (Figure~2) show a point-like source and an extended structure, which may or may not  be related to the point source. Given the large density of background galaxies in the deep \emph{HST} images, it is possible that we are seeing the superposition of a star (either in Draco or in the Galaxy) and a background galaxy/AGN. This could explain why the automated classification is confused (split between AGN and CV). We found no variability between five {\sl XMM-Newton} exposures.

{\bf Source \#16:} This source is strongly (intrinsically) absorbed with an odd jump $>8$ keV in the X-ray spectrum. The source's optical/NIR properties are consistent with those of an evolved low-mass single star, but the apparently high absorption in X-rays rules out an active corona in a nearby star as a source of X-rays. It is possible that this source is a symbiotic star or CV in Draco. In fact, the X-ray spectrum shows signs of emission line(s) from the iron complex (Figure~6). The spectrum also shows hints of other lines at lower (1$-$2\,keV) energies. This could be indicative of a magnetic CV \citep{1997ApJ...474..774F}. We would like to note that symbiotic stars are not a separate class in our current training dataset for MUWCLASS and therefore such a system would be classified as another object type. This source is classified as AGN by the automated algorithm but with low confidence (71\%). The X-ray lightcurve does not exhibit any significant variability.

{\bf Source \#17 :} This source has been reported as a candidate symbiotic star \citep{2000A&AS..146..407B}. \citet{1982ApJ...254..507A} identified this source as carbon star with an unusual SED showing strong emission lines. The authors suggest that the optical colors can be explained by a symbiotic binary (red giant with hot main sequence companion). \citet{1982ApJ...254..507A} also claim that  \#17 is in Draco based on radial velocity measurements. However, the observed X-ray flux implies a high X-ray luminosity of $L_X=1.2\times10^{33}$\,erg\,s$^{-1}$ at the distance of Draco, which can hardly be produced in a non-degenerate binary with red giant and late-type main-sequence star. Therefore, we consider this source to be a good candidate for a quiescent XRB in Draco. 
The automated classification did not produce a confident result. The confusion could be due to the unusual SED noticed by \citet{1982ApJ...254..507A} who also suggested that the star can be in a binary with a (pre-)degenerate object or it can be a star caught very early in the process of ejecting the outer layers and forming a planetary nebula. If the object does not fit any of the predefined object classes the automated algorithm is expected to be confused.

{\bf Source \#19:} Given its X-ray flux and lack of optical counterpart (arguing against binary nature), it is unlikely that this source would belong to Draco (X-ray luminosity, $L_X=4.6\times10^{35}$~erg~s$^{-1}$, for $d=80$~kpc, is too high for isolated NSs), so it could be an old recycled or non-recycled pulsar in a non-accreting binary in our Galaxy. However, the source is imaged strongly off-axis in the \emph{XMM-Newton} image which increases the positional uncertainty, even though source \#19 has one of the smallest positional uncertainties in Table~3 (possibly underestimated by the automated source detection tool in 3XMM). There is an SDSS source within $3''$ with $u=20.6$, $g=20.4$, $r=20.3$, $i=20.3$, and $z=20.3$. If this SDSS source is a counterpart of the X-ray source, it would exclude the possibility of \#19 being nearby isolated pulsar/NS, which otherwise would be possible based on X-ray spectrum and lack of optical/IR counterparts. The lack of optical/IR data prevents our algorithm from confidently classifying this source.

{\bf Source \#23:} The source is faint with a soft X-ray spectrum, too soft for typical AGN. There is some evidence of even further hardening beyond 5$-$6 keV in pn. PL and {\tt bremss} provide a reasonable fit to the data. Our automated algorithm does not provide a confident classification, likely due to incomplete MW parameters (a very faint counterpart found in SDSS, but not in 2MASS). 

Note that nine sources lack counterparts in SDSS. These could be AGNs that are too faint to be detected in SDSS  , or unusual objects that produce little emission in the optical, such as isolated NSs or quiescent BH binaries in Draco. Of those nine, only source \#25 was imaged by \emph{HST} (Figure~2). Two point-like objects are seen within the $r=2''$ circle, one of which could be the counterpart of X-ray source. If this is the case, it could be LMXB/CV in Draco (based on the optical brightness). Unfortunately, no multi-band photometry exist for this field, which prevents us from classifying the possible optical counterpart of the source.  

After careful examination of all X-ray sources we find 4 potentially interesting sources (\#10, \#16, \#17 and \#25) that could belong to Draco and be potential LMXBs, CVs and/or symbiotic candidates. We note in passing that Sculptor is reported to host 5 LMXBs \citep{2005MNRAS.364L..61M} and Fornax to have 2$-$3 ``field'' LMXBs (i.e., not part of Fornax globular clusters, which have another 2 LMXBs; \citealt{2013A&A...550A..18N}). Given the small number statistics and lack of certainty in association with Draco (except, perhaps, source \#16 and \#17), our findings are roughly on par with those for Sculptor and Fornax (both of which are Milky Way satellites and at roughly the same distance as Draco). However, we caution the reader that the respective analyses use different techniques and rely on different MW data and therefore a direct comparison is speculative at best.
 
\citet{2015MNRAS.448.2717W} recently presented optical spectroscopy for stars in the Draco field. They fit stellar models to the all sources and produced a dataset with the derived properties (line-off-sight velocity, $V_{los}$, effective temperature, $T_{\rm eff}$, surface gravity, $g$, metallicity, [Fe/H], etc.). We cross-match the spectroscopic dataset with the positions of the 35 X-ray sources, and found 5 matches -- sources \#9, \#16, \#17, \#22, and \#30. The optical spectroscopy for sources \#9, \#22, and \#30 has very low signal-to-noise (S/N $<3$), which resulted in very uncertain fit parameters and, hence, no classifications are provided in \citet{2015MNRAS.448.2717W}. These low signal-to-noise sources were classified (by our algorithm) as AGN (in agreement with Figure~4). Sources \#16 and \#17 have $V_{los}=-295\pm1$, which is consistent with $V_{los}$ of Draco, and therefore, suggests the association of the two sources with the dSph. Spectroscopy suggests that these two sources are evolved, low-mass stars, which is consistent with CV/LMXB scenarios discussed above.

\subsection{Limits on central IMBH mass}

It has been speculated that dSphs can host an intermediate mass BHs (IMBHs; \citealt{2013MmSAI..84..645N,2005MNRAS.364L..61M} and references therein). We estimate the X-ray luminosity for a putative IMBH in Draco due to Bondi-Hoyle accretion as

\begin{equation}
\begin{split}
L_X\simeq  3.4\times10^{37} \epsilon \left(\frac{M_{BH}}{10^3 ~{\rm M}_{\odot}}\right)^2 \left(\frac{c_s}{10~{\rm km~s^{-1}}}\right)^{-3} \\
\times\left(\frac{n}{0.1 ~{\rm cm^{-3}}}\right)~{\rm erg~s^{-1}}~,
\end{split}
\end{equation}

\noindent where $\epsilon$ is the radiative efficiency, and $n$ is the hydrogen number density in Draco. Based on observational results \citep{1998ARA&A..36..435M,2009ApJ...696..385G}, we estimate an upper limit of $n<0.02$~cm$^{-3}$, and sound speed  $c_s\approx10$~km~s$^{-1}$ for dSphs, which should be of the order of the stellar velocity dispersion \citep{2003AJ....125.1926G,1996MNRAS.282..305H}. Therefore the corresponding flux at the distance of Draco is

\begin{equation}
\begin{split}
F_X\simeq  8.8\times10^{-12}\epsilon\left(\frac{M_{BH}}{10^3 ~{\rm M}_{\odot}}\right)^2 \left(\frac{c_s}{10~{\rm km~s^{-1}}}\right)^{-3} \\
\times\left(\frac{n}{0.02 ~{\rm cm^{-3}}}\right)~{\rm erg~s^{-1}~cm^{-2}}~.
\end{split}
\end{equation}

\noindent  According to the \emph{3XMM} catalog there are four X-ray sources with $S/N\approx3-5$ within $\sim3'$ of the center of Draco ($R.A.=17^h20^m13.2^s$ and $Decl.=+57^{\circ}54'55\farcs3$; \citealt{2003ApJS..145..245R}). The measured flux for the nearest source is $F_X\approx4.5\times10^{-15}$\,erg\,s$^{-1}$~cm$^{-2}$ (the other three sources have similar fluxes and S/N). According to  \citet{2005A&A...440..223B}  the expected radiative efficiency  for synchrotron radiation (presumably the dominate radiation mechanism for a slowly accreting BH) is $\epsilon\sim10^{-5}$. Assuming that one of these X-ray sources is the putative IMBH, we can then estimate its mass. However, it is possible that none of these faint X-ray sources represent the putative IMBH. Therefore, we also can determine the lower limit on the IMBH flux by measuring the diffuse flux fluctuations ($\sim10^{-15}$~erg~s$^{-1}$~cm$^{-2}$ at 3$\sigma$ level) in the local background in the combined EPIC image. This leads to the following constraint 

\begin{equation}
\begin{split}
M_{BH}\left(\frac{n}{0.02~{\rm cm^{-3}}}\right)^{1/2}<3.4\times10^3\left(\frac{F_X}{10^{-15}~{\rm cgs}}\right)^{1/2} \\
\left(\frac{\epsilon}{10^{-5}}\right)^{-1/2}\left(\frac{c_s}{10~{\rm km~s^{-1}}}\right)^{3/2}{\rm ~M_\odot}.
\end{split}
\end{equation}

Theoretical models and observational constraints (e.g., \citealt{2000ApJ...539L..13G}) show that the central BH mass would depend on the velocity dispersion as

\begin{equation}
M_{BH}=1.6\times10^3\left(\frac{\sigma}{10~{\rm km~s^{-1}}}\right)^{3.75}{\rm ~M_\odot},
\end{equation}

\noindent based on the Draco velocity dispersion measurement by Hargreaves et al. (1996). This mass estimate is comparable with the limit calculated above. We note, however, that the mass estimate is based on the use of the upper limit on the hydrogen density and a rather optimistic efficiency of 10$^{-5}$. More conservative values for the density and (especially) the efficiency (where the uncertainty is the largest) could easily reduce the mass limit by several orders of magnitude. 

\subsection{DM decay feature at 3.5 keV?}

We also searched for the $\sim3.5$~keV line that has been previously reported in the Perseus galaxy cluster \citep{2014PhRvL.113y1301B} and M31 \citep{2014ApJ...789...13B}. Figure~7 shows the X-ray spectrum of the diffuse EPIC-pn emission from 0.5 to 10.0 keV. No significant spectral features consistent with previously reported line at $\sim3.5$~keV is seen. 

%
\begin{figure*}
\centering
\includegraphics[width=8.5cm, angle =0 ]{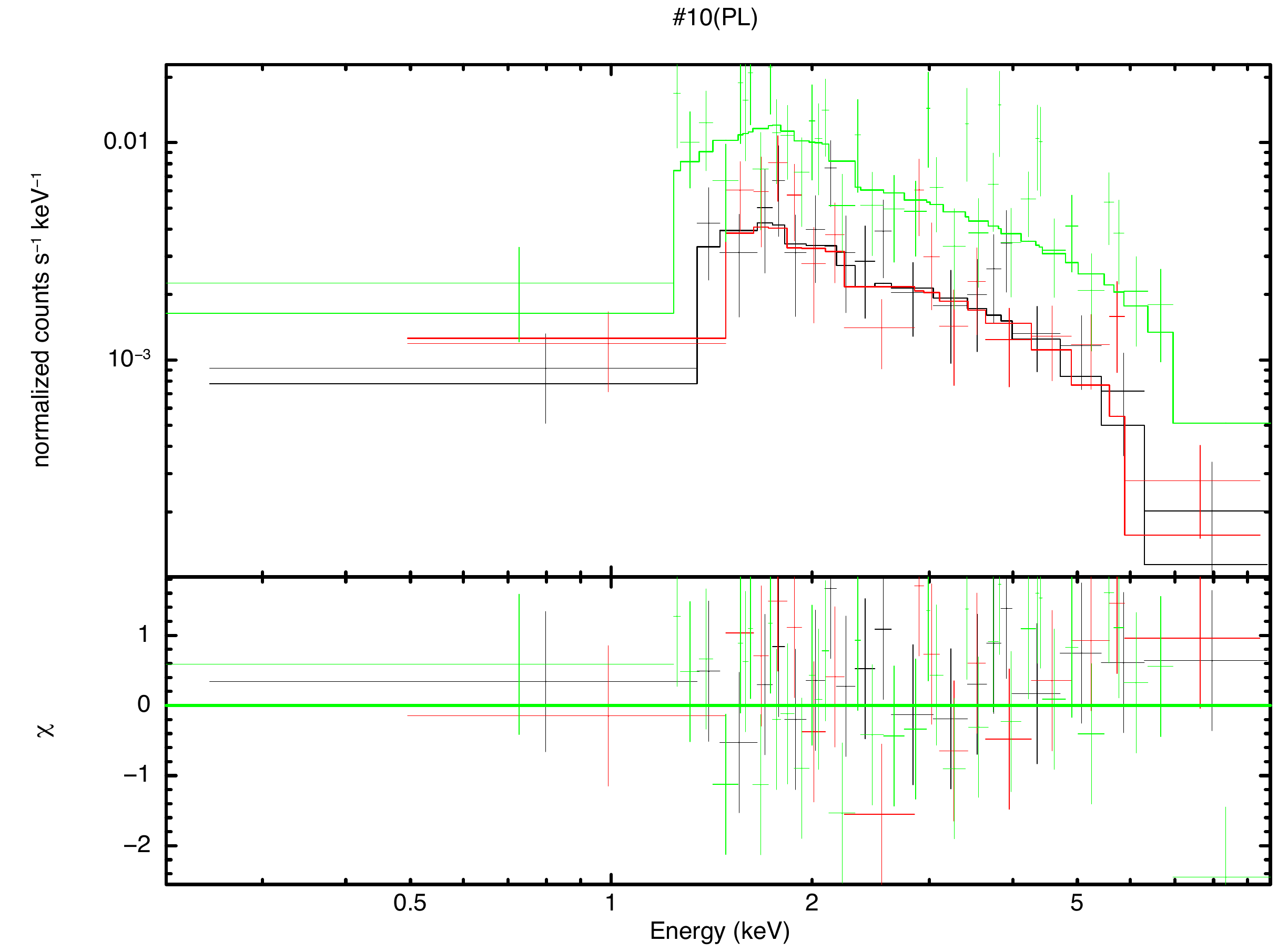}
\vspace{0.1cm}
\includegraphics[width=8.5cm, angle =0 ]{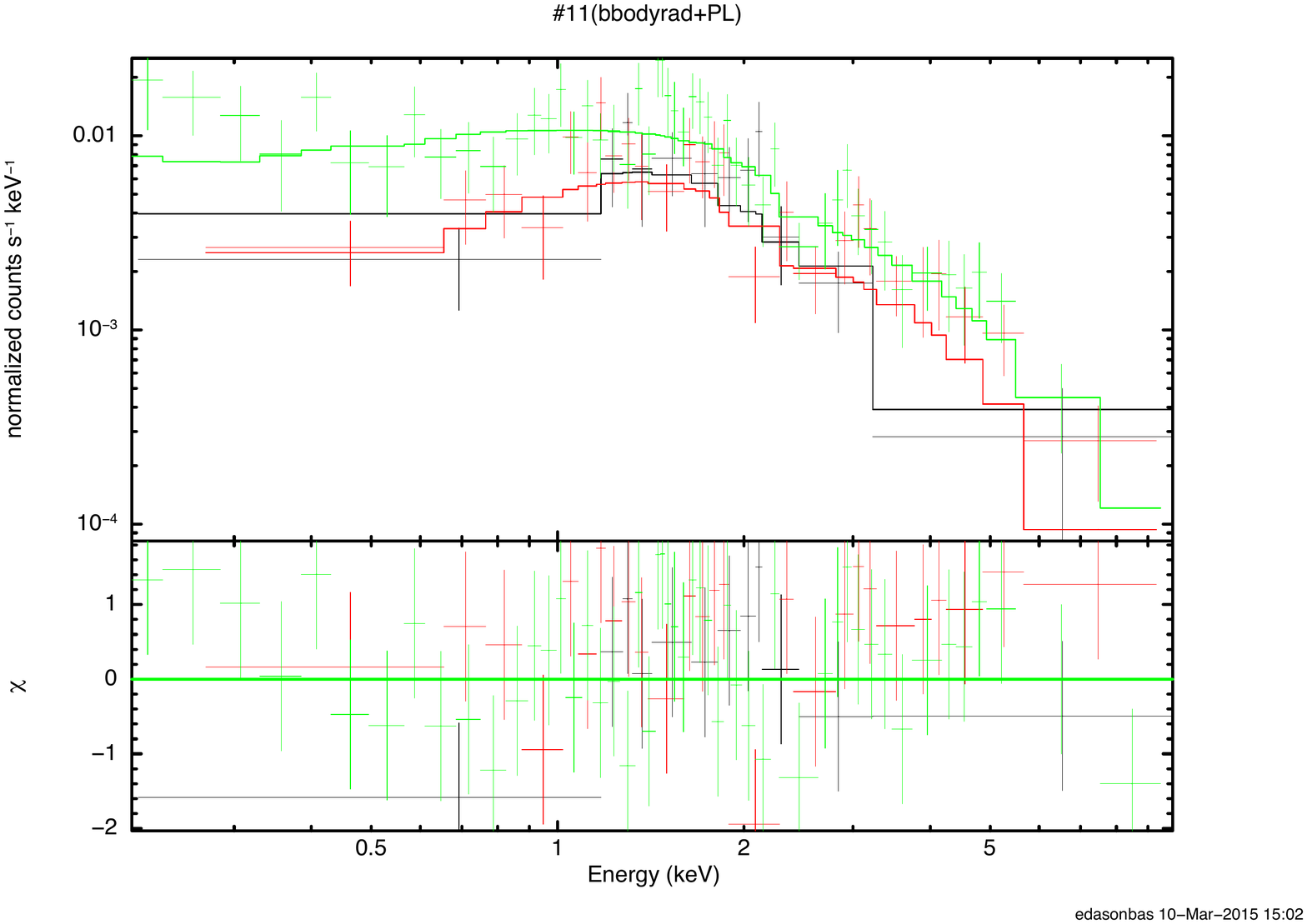}
\vspace{0.1cm}
\includegraphics[width=8.5cm, angle =0 ]{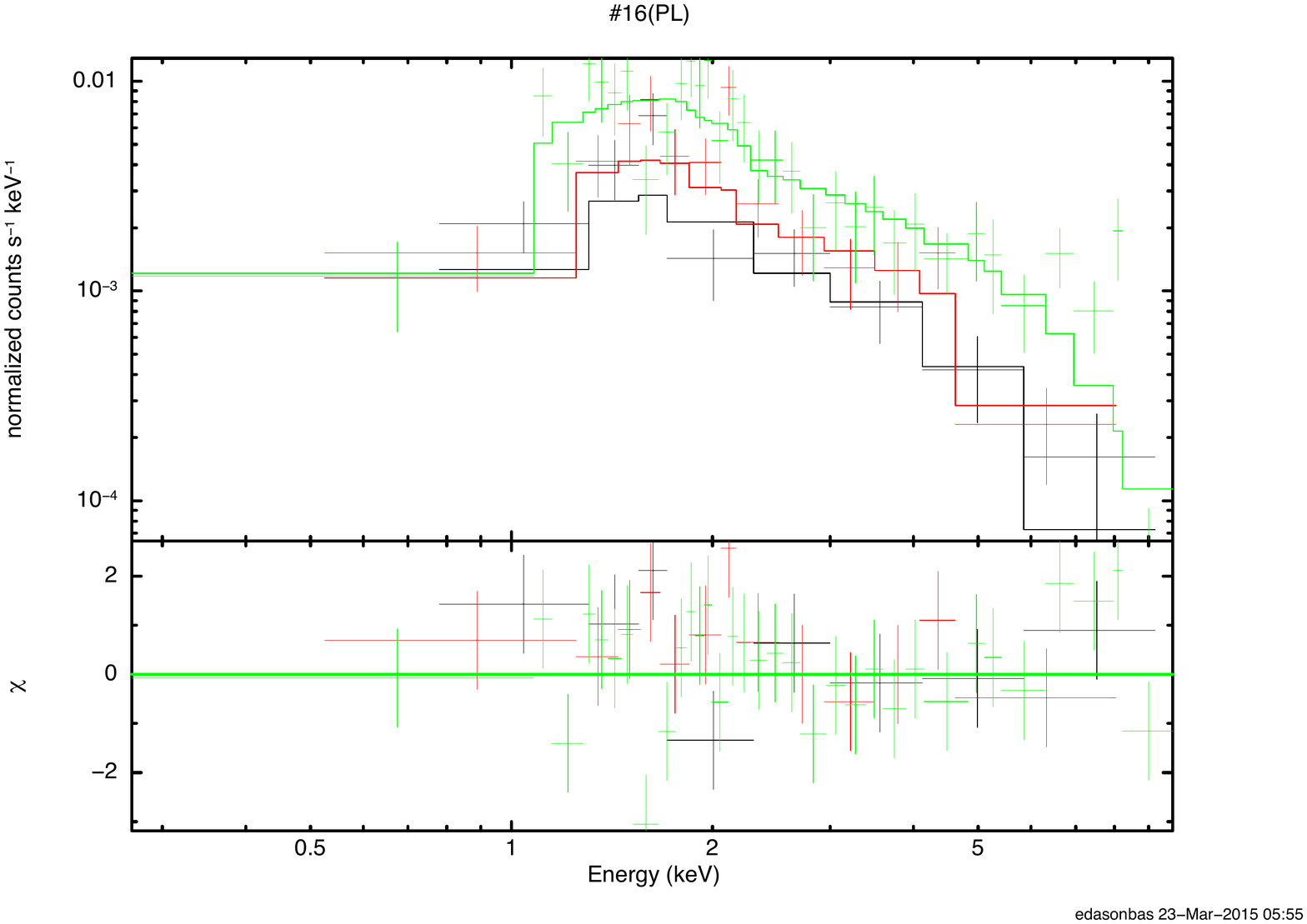}
\vspace{0.1cm}
\includegraphics[width=8.5cm, angle =0 ]{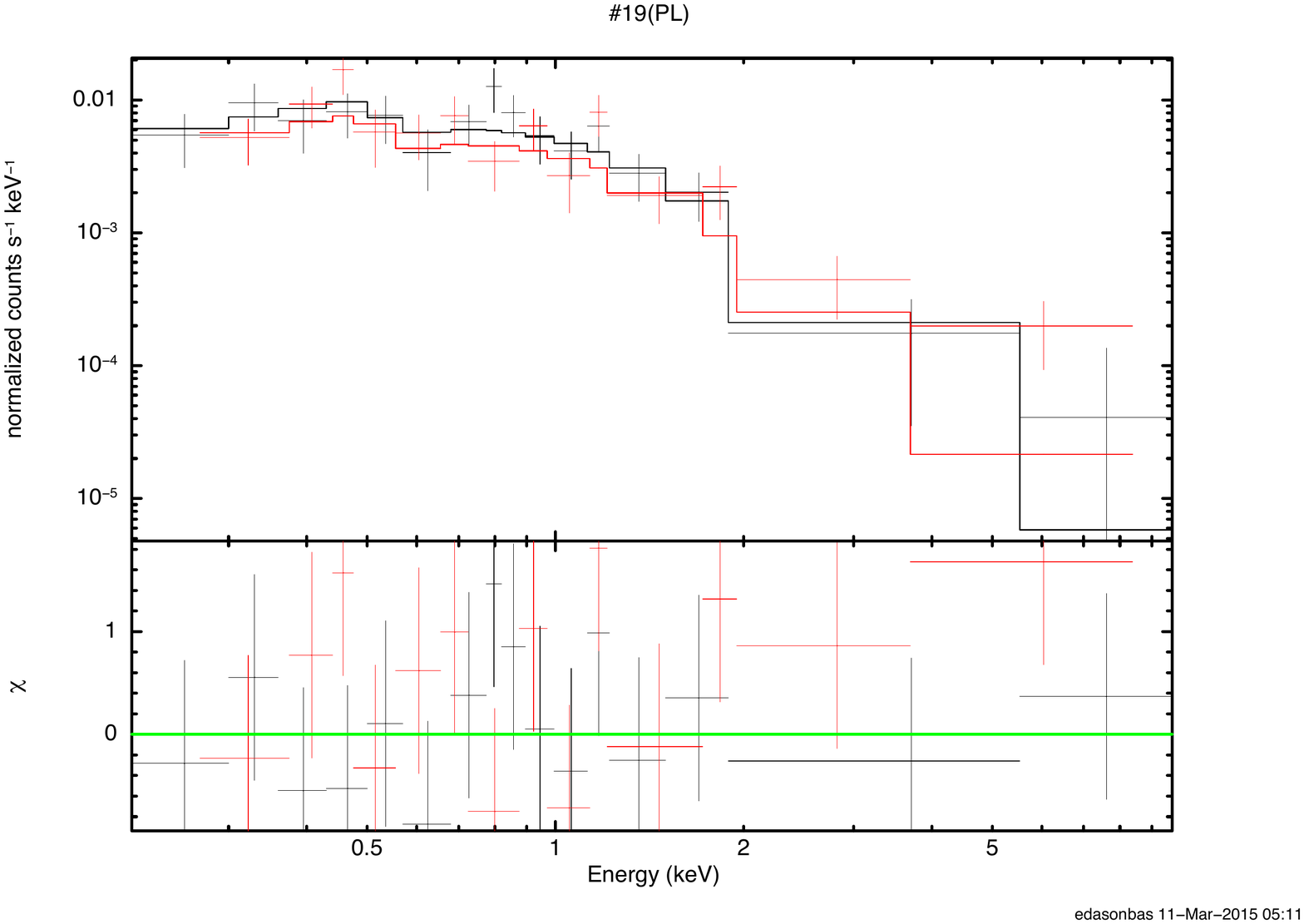}
\vspace{0.1cm}
\includegraphics[width=8.5cm, angle =0 ]{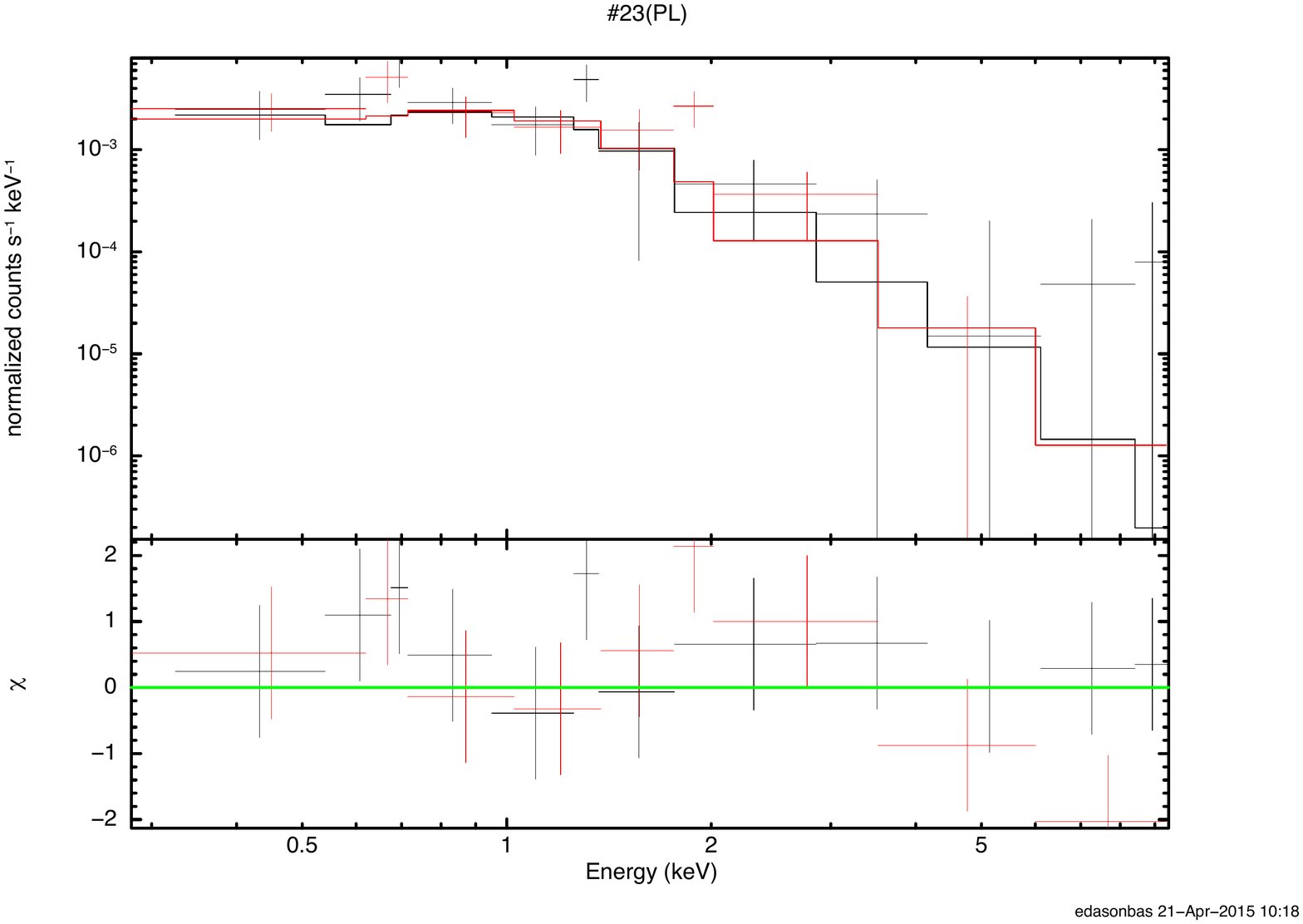}
\vspace{0.1cm}
\includegraphics[width=8.5cm, angle =0 ]{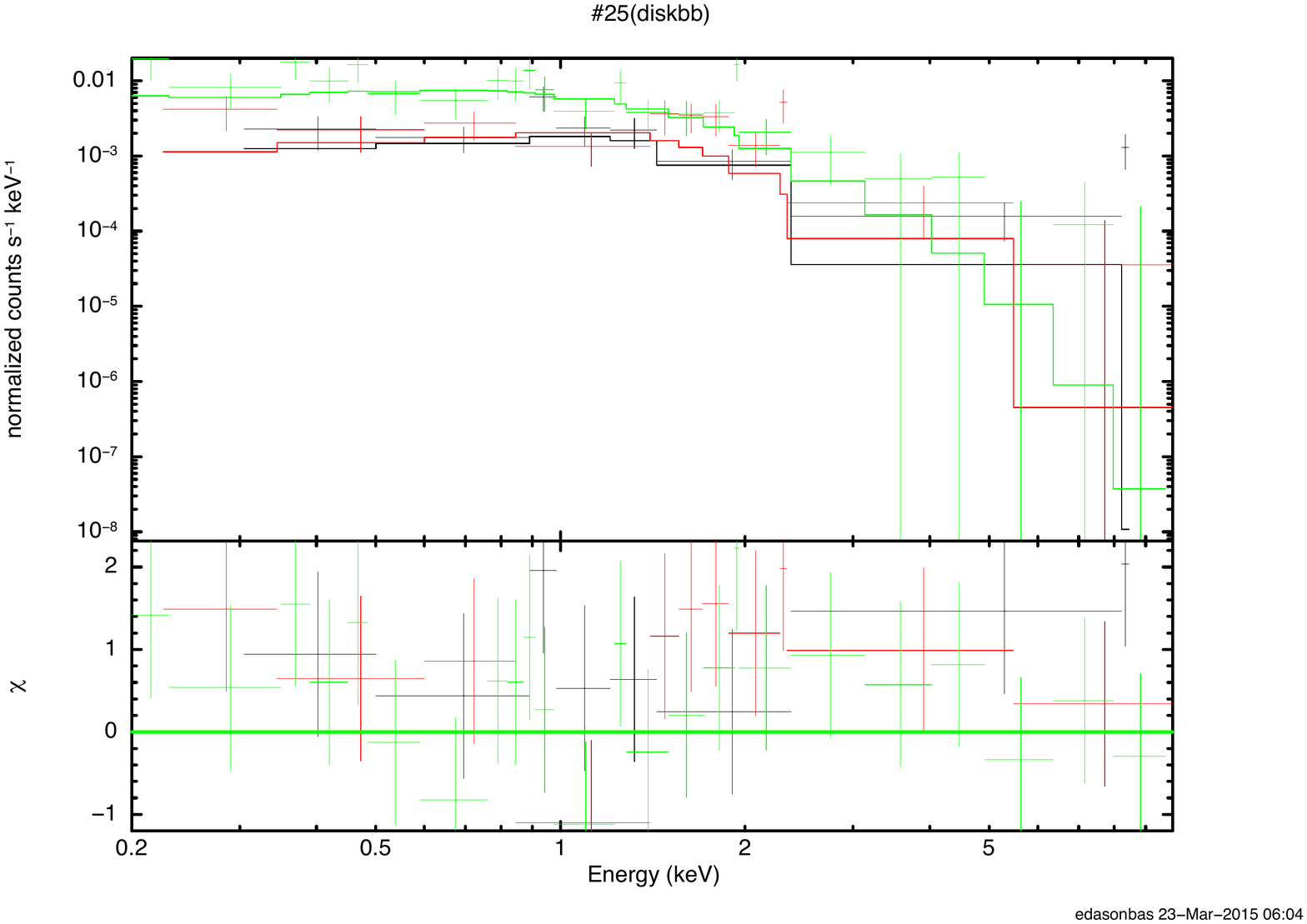}
\caption{Best fit model spectra and their residuals for sources \#10, \#11, \#16, \#19, \#23, \#25 which are not classified as AGN or foreground galactic stars by our automated classification pipeline. MOS1, MOS2, and EPIN-PN data points and their respective best model fits are shown in black, red and green.}
\end{figure*}

\begin{figure}
\begin{center}
\includegraphics[scale=0.3,trim=0 70 0 0]{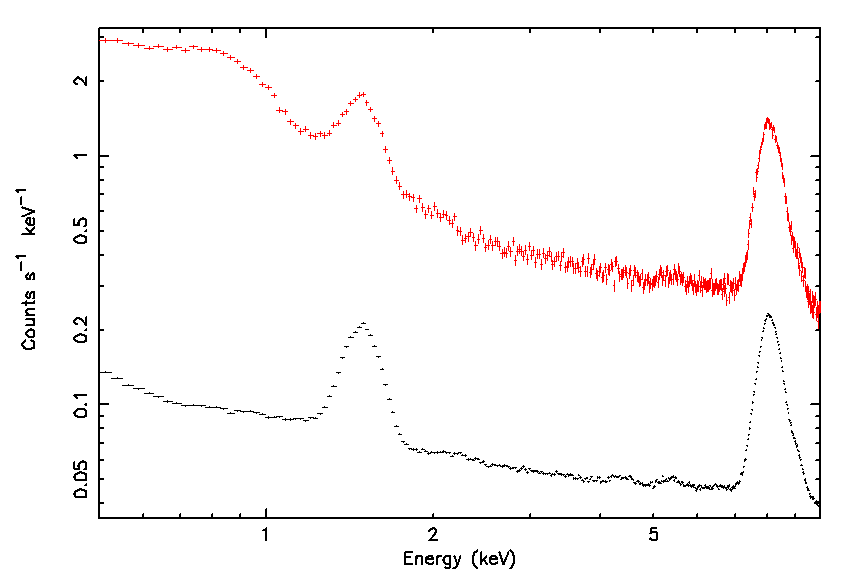}
\vspace{0.4cm}
\caption{The combined (all five observations) EPIC-pn spectrum of the diffuse X-ray emission. The background spectrum for Draco is shown in red in the 0.5-10.0 keV energy range. The filter wheel closed spectrum for the {\sl XMM-Newton} EPIC-pn camera is shown in black. The broad line features are known to be due to instrumental lines; Al K$_\alpha$ at 1.49 keV and Cu florescent at 8 keV. No evidence is found for a spectral feature at $\sim3.5$~keV.}
\end{center}
\end{figure}

\section{Summary}

We have analyzed 35 (bright) X-ray sources detected by \emph{XMM-Newton} in the Draco dwarf spheroidal galaxy. Along with traditional techniques of classification such as the use of color-color and color-magnitude diagrams and hardness ratios, we have utilized an automated machine-learning approach to identify X-rays sources in Draco. These were complimented by the analysis of {\sl HST}/SDSS images and X-ray spectra for several interesting X-ray sources. Our main results are:

\begin{itemize}

\item The classification of X-ray sources in Draco resulted in 12 AGN and 3 foreground stars. We also identified 4 X-ray sources (potential quiescent LMXBs or CVs) that could belong to Draco. For two of them (\#16 and \#17) the associations are supported by the line-of-sight velocity measurements reported in \citet{2015MNRAS.448.2717W}. 

\item The upper  limit on the mass of the IMBH  that we obtained from the X-ray data analysis is similar to the  predictions  based  on  the  velocity  dispersion correlation. The current data are consistent with the none existence of a central black hole, however a  deeper  observation  may  result in such a detection.

\item We do not find any significant evidence of 3.5~keV emission line which could be a signature of decaying DM.

\end{itemize}

Deep multi-band optical observations are required for a more detailed study of the X-ray source population in Draco. Currently, we have 9 sources that lack optical/NIR detections and may represent solitary NSs or quiescent BH binaries in Draco or extended Galactic halo. Combined high-resolution multi-band optical observations, sensitive optical spectroscopy, and a deep \emph{XMM-Newton} observation can provide a much better view into the population of X-ray sources in Draco,  mass of the putative IMBH, and the origin and evolution of dSphs. 

\acknowledgements

ES acknowledges partial funding for this project, provided by the GW Institute for Nuclear Studies, via a summer visitors program and The Science Academy (Bilim Akademisi, Turkey) under the BAGEP program. Partial support for this work was provided by the award NNX09AP84G and is greatfully acknowledged. OK acknowledges support by the National Aeronautics and Space Administration through Chandra Award AR3-14017X issued by the Chandra X-ray Observatory Center, which is operated by the Smithsonian Astrophysical Observatory for and on behalf of the National Aeronautics Space Administration under contract NAS8-03060. 

\appendix{A1. Machine-learning approach to X-ray source classification}\\

Our automated classification tool, MUWCLASS (Brehm et al. 2014; Hare et al. in prep.), uses C5.0\footnote{http://www.rulequest.com}, which is a supervised decision tree learning algorithm requiring a training dataset with already classified sources (see below). C5.0 is an updated version of C4.5 \citep{1993cpml.book.....Q}, which is based on the ID3 algorithm created by  \cite{1986...Q}. We ran C5.0 classification algorithm with the default parameters as described in https://www.rulequest.com/see5-info.html. We have adopted a Laplace prescription for the estimation of the classification confidences \citep{Chawla2003} in each leaf, $P=(TP+1)/(TP+FP+C)$, where $TP$, $FP$, and $C$ are true positives, false positives, and number of classes, respectively, for the leaf where the source in question has ended up. Currently, MUWCLASS does not take into account the  uncertainties of the MW parameters while performing the classification. Our machine-learning classification approach in this paper closely follows that of Lo et al.\ (2014)  and Farrell et al.\ (2015) except that we do not use the X-ray variability parameters since  none of the Draco sources appears to be variable at the statistically significant level.

The training dataset is used to evaluate the parameters of objects from known classes and build the decision tree. The decision tree, calculated from the training data set, is applied to a set of unclassified objects. We have invested significant effort to investigate the catalogs and related literature to create a training dataset of $\approx8,000$ confidently classified X-ray sources detected by either \emph{Chandra} or \emph{XMM-Newton} (Hare et al. in prep.). This dataset includes different numbers of sources from various classes. To compensate for the imbalance we use the SMOTE technique \citep{2011arXiv1106.1813C} to create an expanded balanced dataset. The dataset we used here has 9 predefined objects classes: {\sl (1) main sequence stars} (General Catalog of Variable Stars; Samus et al. 2009), {\sl (2) young stellar objects} (Chandra Orion Ultradeep Point Source Catalog and PAN-Carina; Getman et al. 2005, Povich et al. 2011), {\sl (3) AGNs} (Veron Catalog of Quasars \& AGN; V{\'e}ron-Cetty \& V{\'e}ron 2010), {\sl (4) LMXBs} (Low-Mass X-Ray Binary Catalog, 2007; Liu et al. 2007), {\sl (5) HMXBs} (Catalog of High-Mass X-Ray Binaries in the Galaxy; Liu et al. 2006), {\sl (6) cataclysmic variables} (CVs; Cataclysmic Variables Catalog, 2006, Downes et al. 2001), {\sl (7) isolated neutron stars} (NSs; ATNF Pulsar Catalog; Manchester et al. 2005), {\sl (8) binary non-accreting NS} (ATNF Pulsar Catalog), and {\sl (9) Wolf-Rayet stars} (The VIIth Catalog of Galactic Wolf-Rayet Stars; van der Hucht 2001). We cross-match all objects from the training dataset with MW data and extract the following parameters: X-ray fluxes in 4 bands from the 3XMM-DR5 catalog (the same as those defined in Section 3.2), optical $ugriz$ magnitudes form SDSS, NIR $jhk$ magnitudes from the Two Micron All-Sky Survey (2MASS), and IR W1, W2 and W3 magnitudes from \emph{Wide-field Infrared Survey Explorer} (\emph{WISE}). 

Figure~8 shows the cross-validation matrix for the SMOTEed training dataset which  has the true object class on X-axis and the inferred class on Y-axis. The more diagonal is the matrix, the better is the performance of the classification algorithm. We have also verified the classification performance by dividing the dataset in two parts and using the first part for the classification tree construction and second half for the validation. The dataset is divided into parts before the SMOTE procedure is applied to the part that is used for training. We typically find that $\approx93$\% of the sources are classified correctly, with the AGN and stars being found more confidently than the other sources.

\begin{figure}
\begin{center}
\includegraphics[scale=0.3]{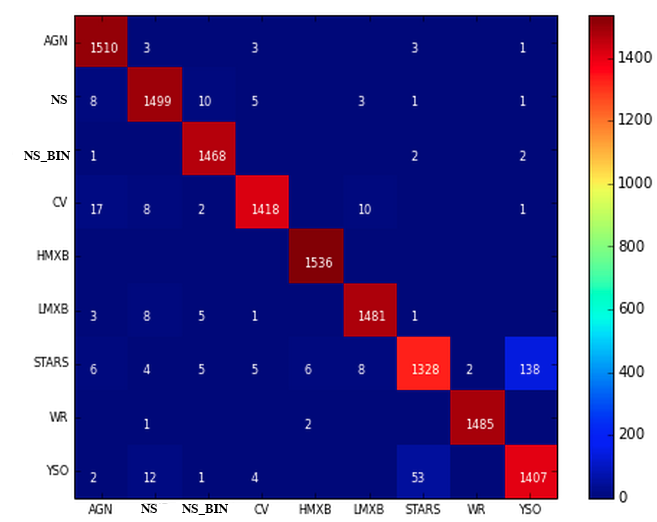}
\vspace{0.4cm}
\caption{The cross-validation matrix of the training dataset (see A1). Original classes are given on the X-axis, and the classification outcome is on the Y-axis. NS\_BIN refers to binary non-accreting NS.}
\end{center}
\end{figure}

\end{document}